\documentclass[final,3p,times]{elsarticle}

\usepackage[utf8]{inputenc}
\usepackage{CJKutf8}
\usepackage{amssymb}
\usepackage{amsmath,amssymb}
\usepackage{algpseudocode}
\usepackage{algorithm}
\usepackage[hyphens]{url}
\biboptions{sort&compress}

\usepackage{graphicx}[dvipdfmx]
\usepackage{comment}
\usepackage{bm}
\newcommand{\vect}[1]{\boldsymbol{\mathbf{#1}}}
\def\vec#1{\vect{#1}}
\usepackage{lineno}
\usepackage{color}
\newcommand\HL[1]{{\color{black}#1}}

\newcommand\HLL[1]{{\color{black}#1}}

\newcommand\HLLL[1]{{\color{black}#1}}

\journal{journal}

\begin{document}

\begin{frontmatter}

\title{Multi-scale physics of cryogenic liquid helium-4: Inverse coarse-graining properties of smoothed particle hydrodynamics}

\author[RCAST]{Satori Tsuzuki~(\begin{CJK}{UTF8}{min}都築怜理\end{CJK})}
\ead{tsuzukisatori@g.ecc.u-tokyo.ac.jp}

\address[RCAST]{Research Center for Advanced Science and Technology, The University of Tokyo, 4-6-1, Komaba, Meguro-ku, Tokyo 153-8904, Japan}

\begin{abstract}
Our recent numerical studies on cryogenic liquid helium-4 highlight key features of multiscale physics that can be captured using the two-fluid model. In this paper, we demonstrated that classical and quantum hydrodynamic two-fluid models are connected via scale transformations: large eddy simulation (LES) filtering links microscopic to macroscopic scales, while inverse scale transformation through SPH connects macro back to microscales. We showed that the spin angular momentum conservation term, introduced as a quantum-like correction, formally corresponds to a subgrid-scale (SGS) model derived from this transformation. \HLLL{Moreover, solving the classical hydrodynamic two-fluid model with SPH appears to reproduce microscopic-scale fluctuations at macroscopic scales.}
\HLLL{In particular,} the amplitude of these fluctuations depends on the kernel radius. \HLLL{This effect may be attributed to truncation errors from kernel smoothing, which can qualitatively resemble such fluctuations; however, this resemblance lacks first-principle justification and should be viewed as a speculative analogy rather than a physically grounded effect.}
Our theoretical analysis further suggests that the Condiff viscosity model can act as an SGS model, incorporating quantum vortex interactions under point-vortex approximation into the two-fluid framework. These findings provide new insight into the microscopic structure of cryogenic helium-4 within a multiscale context. Notably, the normal fluid can be understood as a mixture of inviscid and viscous fluid particles. \HLL{While molecular viscosity renders the normal fluid at microscopic scales, its small magnitude contributes little to the large-scale effective viscosity, which includes both molecular and eddy viscosities; therefore, in laminar regimes where eddy viscosity is negligible, the normal fluid may be effectively treated as inviscid at large scales if molecular viscosity is sufficiently small.}
\end{abstract}


\end{frontmatter}







\section{Introduction}
The unique behavior of liquid helium-4 at cryogenic temperatures---specifically, its loss of viscosity near absolute zero---has long intrigued physicists. Each helium-4 atom, composed of two protons and two neutrons, is a boson, allowing the liquid to be modeled as a quantum many-body system of interacting bosons. This system has been widely studied in condensed matter physics. Research began with the liquefaction of helium by H. K. Onnes~\cite{TF9221800145}, the discovery of superfluidity by P. L. Kapitza~\cite{Kapitza1938}, and its theoretical interpretation via Bose--Einstein condensation by F. W. London~\cite{LONDON1938}. By the mid-20th century, it was established that below the critical temperature ($\approx$ 2.17 K), a fraction of atoms occupy the ground state, with kinetic energy limited to the zero-point vibrational level. Being a noble monatomic gas, helium-4 has stable electron orbitals, minimal polarizability, and weak van der Waals interactions, resulting in exceptionally low viscosity below the critical temperature. Experimentally, E. F. Burton reported an eightfold viscosity drop across the lambda point at high Reynolds number~\cite{BURTON1935}, while Kapitza observed a reduction by a factor of at least 1500 in laminar flow~\cite{Kapitza1938, ALLEN1938}.

Phenomenological studies of liquid helium-4 have progressed alongside microscopic theories. The two-fluid model, introduced by L.Tisza~\cite{TISZA1938} and L. Landau~\cite{PhysRev.60.356}, posits coexistence of a viscous normal component and an inviscid superfluid component at cryogenic temperatures. Their model, based on bosonic statistics, successfully explained counterflow experiments under low heat input and predicted two types of sound waves, later confirmed experimentally~\cite{Atkins01041952, Dingle01041952}. However, it failed to describe heat--velocity relationships at higher heat flux. To address this, Gorter and Mellink~\cite{GORTER1949285} introduced mutual friction between the components, extending the model's applicability. We refer to this extended version as the two-fluid model based on quantum hydrodynamics or the quantum hydrodynamic two-fluid model.
Recent developments interpret counterflow microscopically through interactions between normal fluid and quantum vortices, which appear when fluid velocity exceeds Feynman's critical value~\cite{FEYNMAN195517}, signaling the breakdown of pure superfluidity. Quantum vortices, unlike classical ones, are stable and non-dissipative due to phase quantization, with circulation quantized in units of Planck's constant. Their dynamics and induced velocity fields are described by the Biot--Savart law, and the vortex filament model (VFM) offers a practical framework for solving it~\cite{doi:10.1063/1.5091567, doi:10.1073/pnas.1312535111}. Recent simulations couple the VFM with the Navier--Stokes equations to analyze counterflow~\cite{Idowu2001, 10.1063/1.4828892, PhysRevLett.124.155301}.
Circulation quantization also plays a key role in rotational phenomena. One notable consequence is the persistent current, where a superfluid sustains steady flow indefinitely in a toroidal container~\cite{Kojima1972}. Moreover, under horizontal rotation, helium-4 forms quantum vortex lattices---regular arrays of aligned vortices that rotate rigidly about the container axis, with velocity increasing linearly with the distance from the axis. These phenomena observed in experiments~\cite{Yarmchuk1982, PhysRevLett.86.4443} and simulations~\cite{PhysRevA.65.023603, PhysRevA.67.033610, Adhikari_2021} exemplify the distinct rotational behavior of quantum fluids.

Liquid helium-4 has primarily been studied within the frameworks of quantum mechanics and quantum hydrodynamics, with most numerical models focusing on nanometer-scale phenomena. However, simulating bulk-scale (centimeter to meter) behavior using such models is computationally prohibitive due to the vast number of particles or grid points required, even with modern supercomputers. Macroscopic phenomena such as film flow~\cite{Vinen:808382}---where superfluid helium climbs container walls due to vanishing viscosity---and the fountain effect~\cite{Amigo_2017}---where heating causes helium to flow through a porous medium and eject from the container top---exemplify so-called macroscopic quantum phenomena. Though visible to the naked eye and occurring on classical scales, these effects are governed by quantum-induced vanishing viscosity and thus fall outside traditional fluid mechanics. 

Rotational phenomena offer similar examples. While vortex cores exist on the angstrom scale and their velocity fields extend to microns, the overall vortex lattice structure spans microns to centimeters. For instance, in a circular vessel 1 cm in diameter rotating at 5 $\rm rad \cdot s^{-1}$, the vortex spacing estimated from Feynman's rule~\cite{FEYNMAN195517, VANSCIVER2009247} is $10^{-2}$ cm clearly within the classical hydrodynamic regime. Similarly, persistent current experiments involve toroidal vessels with diameters of several millimeters and circumferences on the order of tens of centimeters~\cite{Kojima1972}. Despite this, no simulation framework currently exists that solves the Navier--Stokes (NS) equations while accurately capturing such macroscopic quantum behavior. Moreover, viscosity in rotational flows behaves more complexly than in shear-dominated cases. In rotational viscometer experiments, viscosity sharply decreases at the superfluid transition, then rises again and diverges as temperature approaches absolute zero~\cite{Hollis-Hallett_1953}. In summary, the fundamental mechanisms behind rotational behavior in superfluid helium remain unresolved, and numerical methods capable of modeling macroscopic quantum phenomena in bulk helium are yet to be developed.

\begin{figure}[t]
\vspace{-0.8cm}
\centerline{\includegraphics[width=1.0\textwidth, clip, bb= 0 0 2060 831 ]{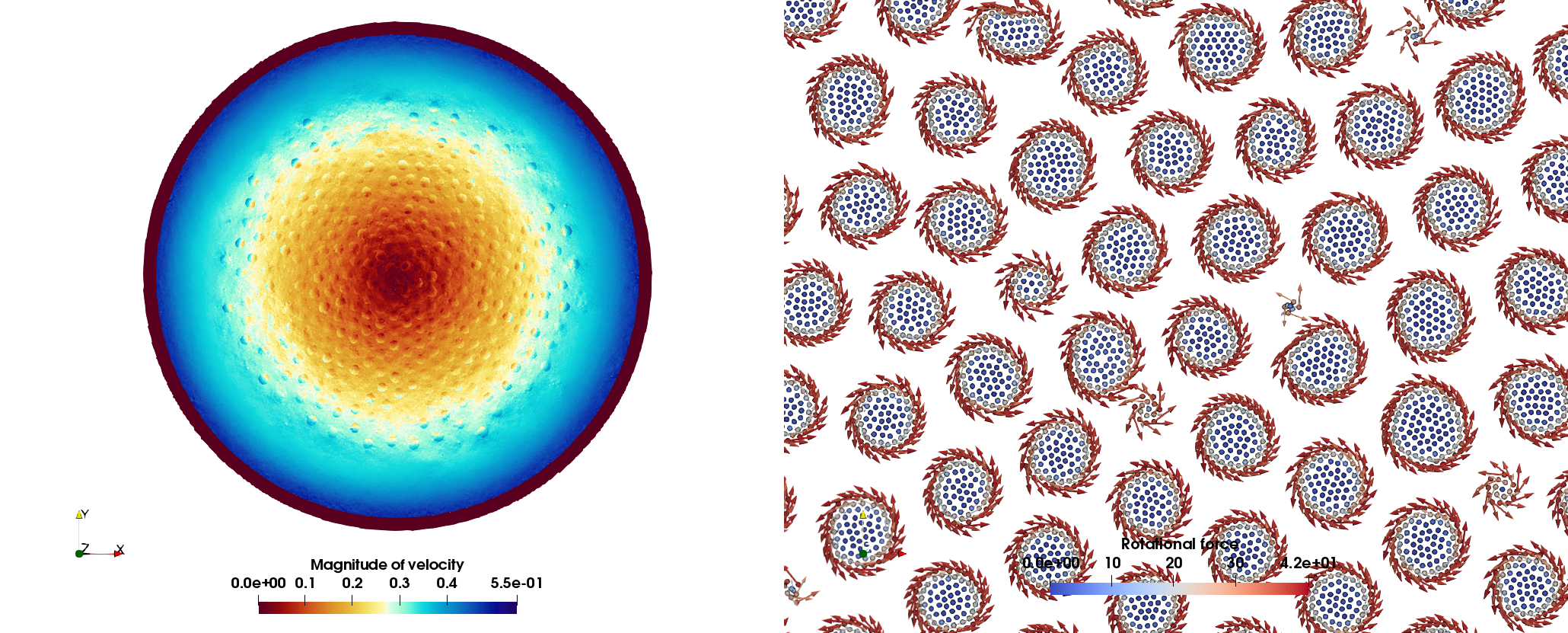}}
\caption{Snapshot of \HLLL{the simulation performed under the same conditions as those reported} in Refs.~\cite{doi:10.1063/5.0060605, doi:10.1063/5.0122247}. (a) shows the intensity distribution of the particle velocity, and (b) is an enlarged view of (a), visualized by the direction and intensity of the rotational force of individual vortices as arrow vectors and color maps, respectively.}
\label{fig:Figure-Video-VortexLattice2D}
\end{figure}

\begin{figure}[t]
\vspace{-2.0cm}
\centerline{\includegraphics[width=0.80\textwidth, clip, bb= 0 0 840 691 ]{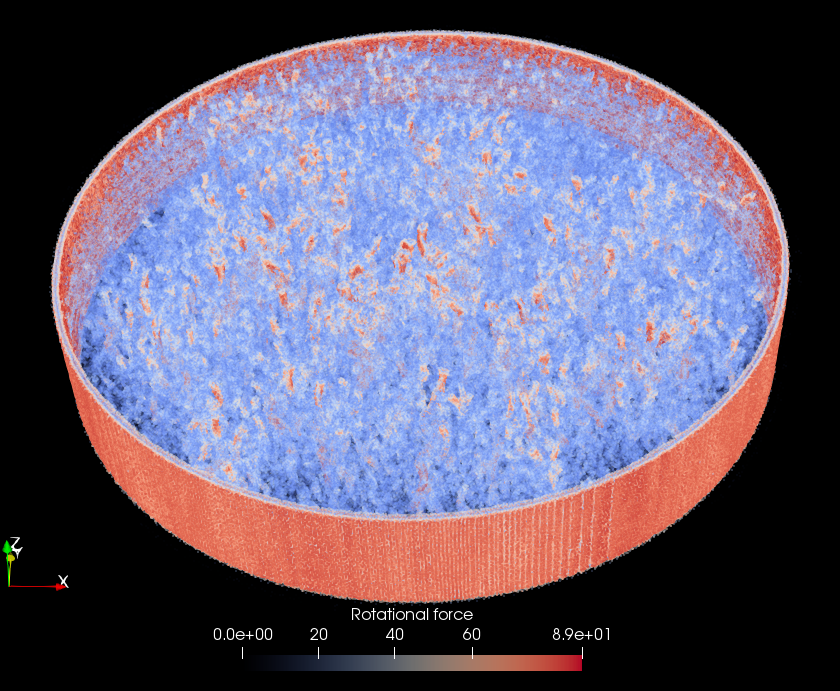}}
\caption{Snapshot of \HLLL{the simulation performed under the same conditions as those reported} in Ref.~\cite{10.1063/5.0218444}; the liquid helium-4 \HLLL{can be observed} rotating only in the horizontal direction with a certain thickness.}
\label{fig:Figure-Video-RotatedHelium3D371}
\end{figure}

\begin{figure}[t]
\vspace{-47.5cm}
\hspace{27.7cm}
\centerline{\includegraphics[width=4.3\textwidth, clip, bb= 0 0 4056 3375 ]{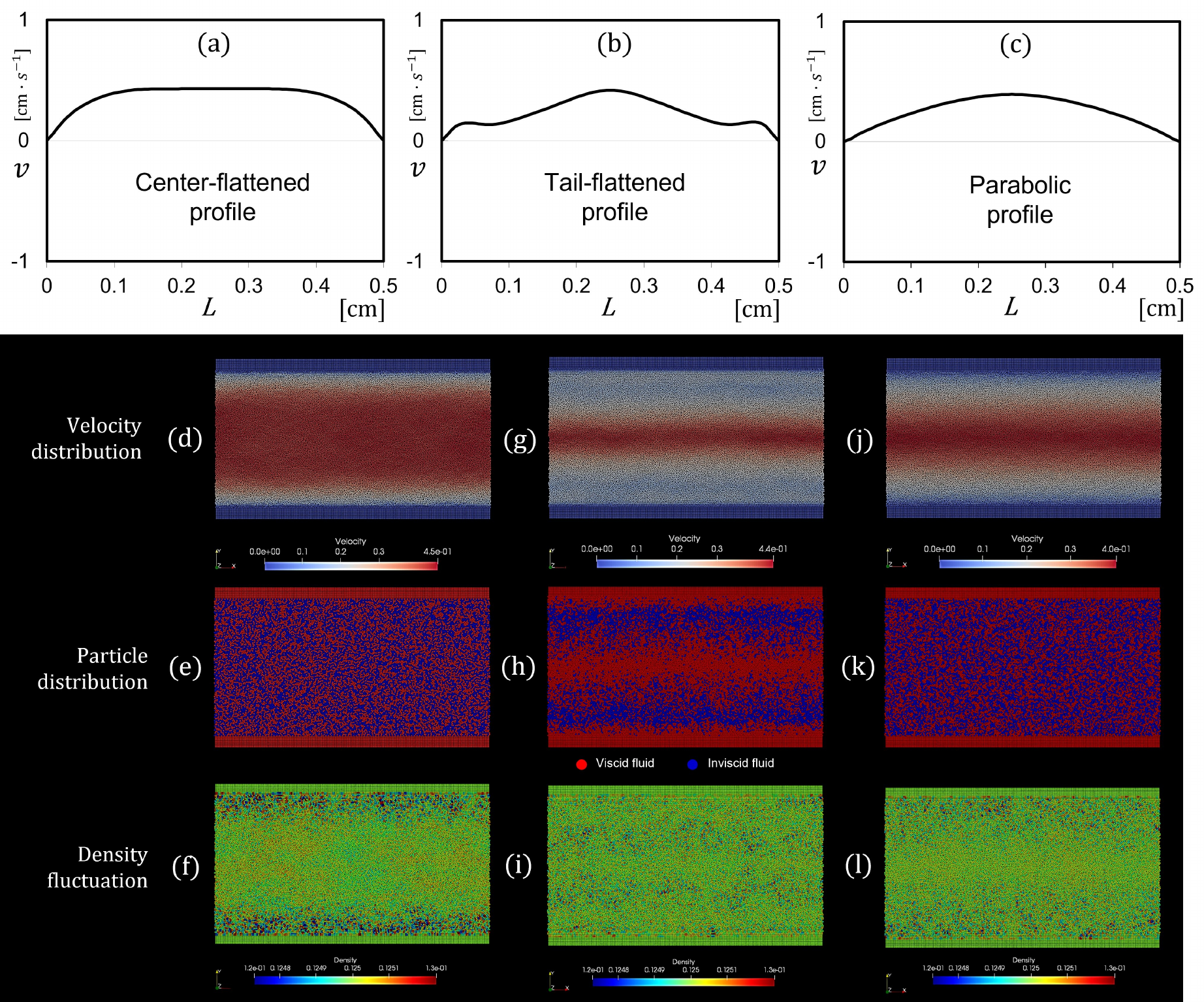}}
\caption{Comparison of the simulation results of the counterflow using our two-fluid model based on classical hydrodynamics, demonstrating that all three representative velocity profiles, (a) center-flattened, (b) tail-flattened, and (c) parabolic, were successfully obtained using only the difference in the initial particle distributions. (d)--(f) Velocity distribution, initial particle distribution, and density fluctuation for the center-flattened profile, (g)--(i) for the tail-flattened profile, and (j)--(l) for the parabolic velocity profile, respectively.}
\label{fig:Figure-ResultsCounterFlow}
\end{figure}

To address the challenge of simulating cryogenic bulk liquid helium-4, we developed a numerical method grounded in classical fluid dynamics, aiming to reproduce macroscopic quantum phenomena. This approach builds on key insights from previous studies. Landau's two-fluid model treated the inviscid and viscous components separately via the Euler and Navier--Stokes (NS) equations. However, at high heat flux, mutual friction between the components becomes significant, indicating that a suitably corrected classical two-phase flow may approximate helium-4 dynamics under specific heated conditions. Here, we distinguish between quantum corrections, which involve Planck's constant and lead to quantized physical quantities (e.g., energy, angular momentum), and particle corrections, which reflect discrete particle behavior beyond the continuum approximation. In earlier work, we pointed out a formal similarity between the quantum hydrodynamic two-fluid model and classical two-phase flow models. Building on this observation, we discretized the governing equations for inviscid and viscous fluids, incorporating interfacial interactions and inter-component forces~\cite{Tsuzuki_2021, doi:10.1063/5.0060605}. The model allows for particle corrections at interfaces and enforces spin angular momentum conservation as a minimal quantum correction. This framework---referred to as the two-fluid model based on classical hydrodynamics, the classical hydrodynamic two-fluid model, or simply our model---is intended to bridge quantum behavior and macroscopic fluid dynamics.

In our previous work, we employed smoothed particle hydrodynamics (SPH)—a Lagrangian method that discretizes the fluid equations by tracking individual particles. Originally developed in astrophysics~\cite{gingold1977smoothed, 1985AA149135M, monaghan1992smoothed}, SPH has since been applied to interfacial phenomena, including free-surface flows~\cite{MONAGHAN1994399} and multiphase systems~\cite{MONAGHAN1995225, MORRIS199741, Monaghan_2005}. This finite-particle approximation represents a fluid continuum as a system of interacting fluid particles. In classical fluids, these particles lack intrinsic meaning and merely approximate the continuum. However, liquid helium-4 at cryogenic temperatures is a bosonic many-body system, allowing each virtual fluid particle to be interpreted analogously to a Bose particle rather than as a phenomenological fragment of a fluid. Thus, fluid particles in this context may be viewed as coarse-grained or representative of quantum particles. While a direct correspondence with microscopic behavior has yet to be firmly established, further investigation into such connections may justify treating these particles as classical virtual fluid elements that exhibit quantum-like features. Notably, formal parallels exist between our finite-particle two-fluid model and quantum many-body systems, particularly in conserving spin angular momentum. Each quantum particle possesses quantized spin. Likewise, SPH enables local conservation of angular momentum around individual particles. For instance, in a two-dimensional (2D) system, assuming particles are rigid, vertically aligned, and uniform, constant angular velocity implies conservation of spin angular momentum. While this rotational motion is a classical analogy and may not reflect quantum kinetics, it mathematically reproduces key features of quantized angular momentum.

The spin angular momentum-conserving Navier--Stokes (NS) equation was originally derived for polar fluids by Condiff et al. in 1964~\cite{doi:10.1063/1.1711295}. Later, M{\"{u}}ller~\cite{MULLER2015301} introduced a discrete SPH formulation of this equation within the smoothed dissipative particle dynamics (SDPD) framework for simulating suspension flows, such as blood flow. Inspired by these developments, we applied a similar formulation to the normal fluid component of Landau's two-fluid model, enabling spin angular momentum conservation for individual fluid particles. This approach successfully reproduced key features of quantum vortex lattices, namely the rigid-body-like rotation of multiple vortices around a vessel's axis, each maintaining its own spin without dissipation~\cite{doi:10.1063/5.0060605, doi:10.1063/5.0122247, Tsuzuki_2021}. Figure~\ref{fig:Figure-Video-VortexLattice2D} shows a replicated simulation based on Refs.~\cite{doi:10.1063/5.0060605, doi:10.1063/5.0122247}. Panel (a) displays particle velocity intensity, revealing distortions at vortex cores. Panel (b) magnifies this region, illustrating the rotational force of each vortex using vector arrows and a color map. Figure~\ref{fig:Figure-Video-RotatedHelium3D371} reproduces results from Ref.~\cite{10.1063/5.0218444}, showing helium-4 rotating horizontally with a finite thickness. The configuration is quasi-two-dimensional and inherently anisotropic. Since vortex interactions in three dimensions must obey topological constraints---e.g., Schwarz's rule~\cite{PhysRevB.31.5782, PhysRevB.38.2398}---the interactions shown here are not fully accurate. Nevertheless, the spontaneous formation of multiple vortices was captured. Future work will focus on extending the model to fully three-dimensional dynamics by incorporating appropriate topological treatments via Schwarz's rule.

In Ref.~\cite{doi:10.1063/5.0122247}, we simulated a 2D counterflow---another hallmark problem in cryogenic helium-4---and reproduced both the classical parabolic velocity profile (Hagen--Poiseuille flow) and a center-flattened profile, where the central velocity flattens. While such flattening also occurs in classical turbulent flows and is not exclusive to quantum fluids, a tail-flattened profile~\HLL{\cite{PhysRevB.91.094503}}---where flow near the walls is suppressed and the central region elevated---is considered unique to quantum hydrodynamics, as confirmed by both experiments and simulations. Its origin remains under debate. One hypothesis attributes it to a laminar normal component coexisting with a turbulent superfluid. More recently, spatial variations in vortex line density have been proposed as a key factor~\cite{Kobayashi2019}: a uniform vortex distribution tends to flatten the center, while wall-localized vortices yield a tail-flattened profile. Because quantum vortices are stable and nondissipative, their interaction with the fluid can be modeled as a fluid--structure interaction, wherein the sfluid experiences drag from spatially distributed vortex structures. In this framework, each coarse-grained particle represents a bundle of quantum vortices (or vortex filaments in 3D), and the drag it exerts is a macroscopic, spatially averaged force derived from many microscopic interactions. 

\begin{figure}[t]
\vspace{-1.5cm}
\centerline{\includegraphics[width=0.85\textwidth, clip, bb= 0 0 1020 716 ]{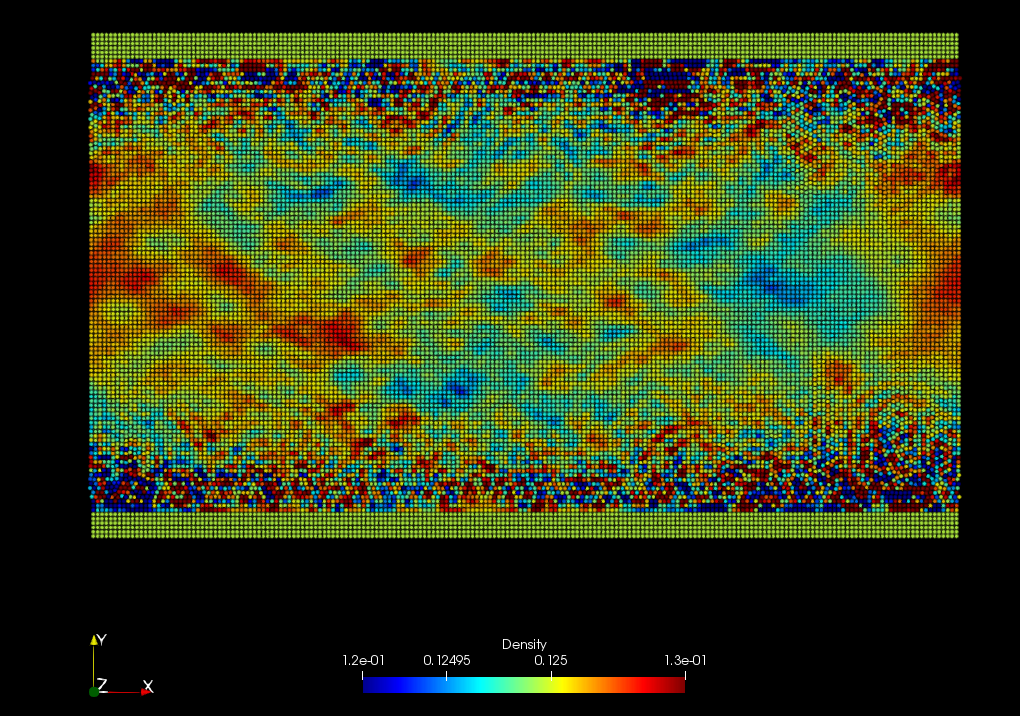}}
\caption{Snapshot of the simulation demonstrating the emergence of shock waves in the explicit SPH calculations, in which a steeper temperature gradient was given to the system, resulting in abrupt and significant density changes occurred in time and space compared with the case in Fig.~\ref{fig:Figure-ResultsCounterFlow}(a). The density oscillation around 1 $\%$, which is not usually emphasized, is intentionally enlarged for clarity.}
\label{fig:Figure-Video-ResultsSoundWave}
\end{figure}

As an extension of our previous work~\cite{doi:10.1063/5.0122247}, we conducted counterflow simulations using our two-fluid model and successfully reproduced a tail-flattened velocity profile, a feature characteristic of quantum fluids. In our model, the residual viscous component---low in density at cryogenic temperatures---forms vortex cores [33]. To emulate the vortex line density distribution described in Ref. [47], we adjusted the initial distribution of the viscous component accordingly: uniform for the parabolic (c) and center-flattened (a) cases, and concentrated near the walls for the tail-flattened (b) case. All three simulations were conducted under the same conditions, except that case (c) included a tenfold temperature gradient. As shown in Figs.~\ref{fig:Figure-ResultsCounterFlow}(a)--(c), we reproduced all three characteristic velocity profiles observed in helium-4 counterflow. Notably, the tail-flattened profile in (b) was achieved solely by altering the spatial distribution of particles within a classical hydrodynamic framework.
Figures~\ref{fig:Figure-ResultsCounterFlow}(d), (g), and (j) show the velocity profiles corresponding to (a)--(c), with (a)--(c) being x-direction averages. Figures~\ref{fig:Figure-ResultsCounterFlow}(e), (h), and (k) show the corresponding distributions of the two fluid components, indicating that the initial spatial particle distributions are retained over time. Figures~\ref{fig:Figure-ResultsCounterFlow}(f), (i), and (l) show density fields, with color maps reflecting variations within $\pm$ 4 $\%$ of the mean. In all cases, velocity tends to decrease in regions of higher density fluctuation. Fluctuations reached ~0.35 $\%$ in (a), and up to ~0.2 $\%$ in (b) and (c), consistent with the stronger thermal gradient in (c). These results demonstrate that our two-fluid model might capture the tail-flattened velocity profile through only fluid particle interactions at fluid dynamics scale. In summary, we successfully reproduced phenomena previously regarded as uniquely quantum, including vortex lattices and counterflow profiles, using a two-fluid model based on classical hydrodynamics.

\begin{figure}[t]
\vspace{-7.0cm}
\hspace{0.6cm}
\centerline{\includegraphics[width=1.1\textwidth, clip, bb= 0 0 1030 831 ]{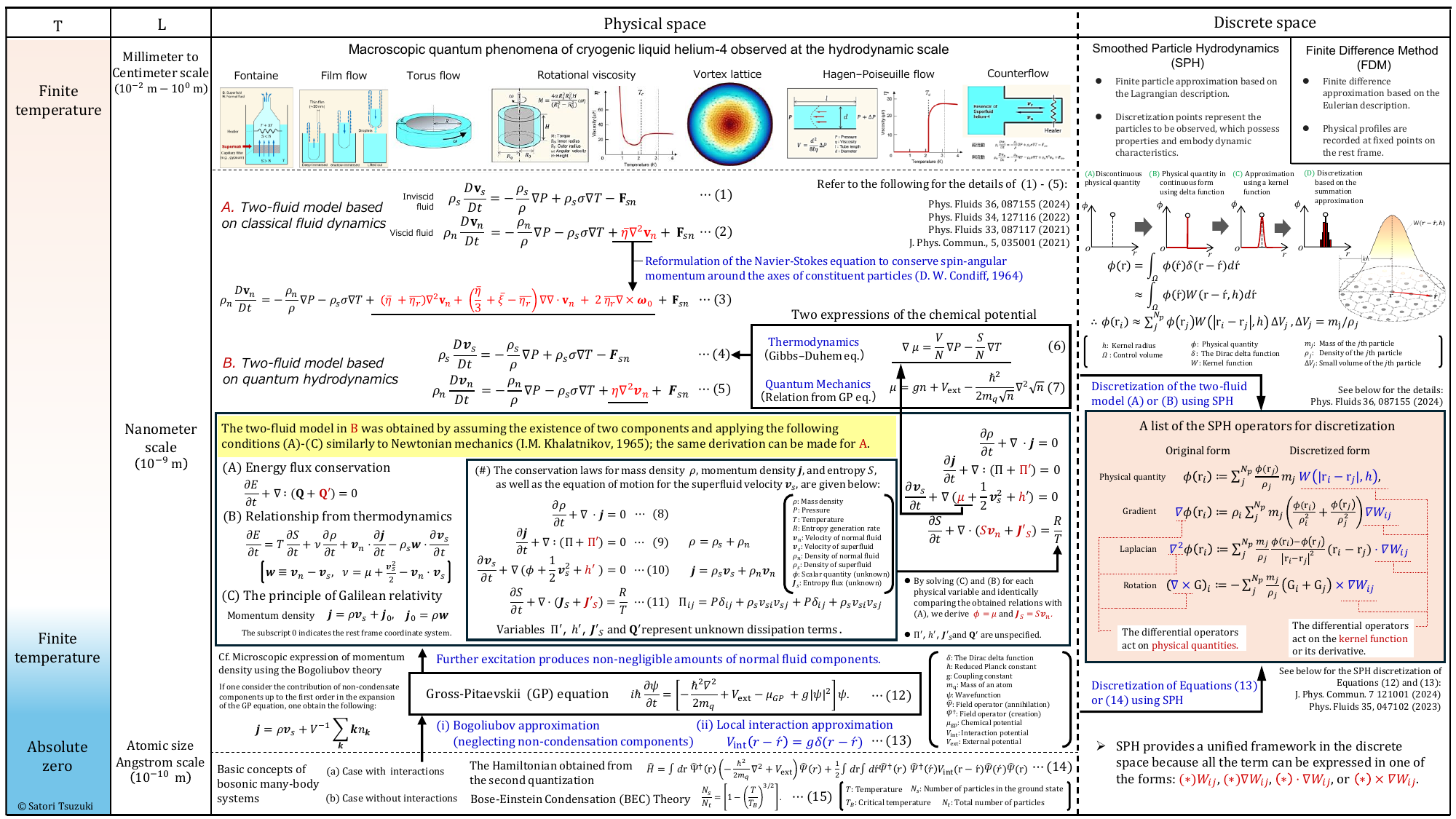}}
\caption{Schematic of the theoretical models representing the dynamics of liquid helium-4, which are closely associated with our two-fluid model based on classical hydrodynamics, classified by temperature and spatial scale in physical and discrete space, summarized on a single page and embedded in vector form for easy enlargement. The correspondence between the fourth and fifth terms on the right side of Fig.~\ref{fig:Figure-ComparisonWithRelated}(3) with the divergence of the subgrid scale stress tensor (-$\nabla \cdot \vec{\tau}_{SGS}$) is presented in Sec.~\ref{sec:closure}. }
\label{fig:Figure-ComparisonWithRelated}
\end{figure}

Furthermore, in SPH simulations of incompressible flows, shock-like behavior can arise from local density fluctuations. When density changes cause fluid velocities to exceed the Courant--Friedrichs--Lewy (CFL) limit~\cite{XU201643}, the continuum assumption fails, generating shocks. Figure~\ref{fig:Figure-Video-ResultsSoundWave} shows such a case with a steeper temperature gradient leading to a 1.1 $\%$ density oscillation and shockwave emergence (see supplemental video). While such behavior is typically unwelcomed in ordinary CFD for incompressible flow, where incompressibility is strictly enforced, it resembles first sound waves in cryogenic helium-4, suggesting SPH may naturally capture such phenomena. These findings raise a key question: Why does the SPH formulation appear to facilitate the ability of our two-fluid model to reproduce macroscopic quantum phenomena that are typically attributed to microscopic effects, such as the loss of molecular viscosity? We address this by situating our model in the broader context of quantum hydrodynamics. Our results indicate that SPH discretization introduces a reverse coarse-graining effect, where kernel-based smoothing errors can emulate microscopic fluctuations at macroscopic scales, with amplitudes scaling with the kernel radius. We further propose that this behavior parallels an inverse scale transformation, akin to the filtering used in large eddy simulations (LES), connecting classical and quantum hydrodynamic two-fluid models. Moreover, the spin angular momentum conservation term---originally introduced as a quantum correction---can be interpreted as a subgrid-scale (SGS) model derived from such scale transformations. Our results and discussion provide new insights into the microscopic composition of cryogenic liquid helium-4 in a multiscale framework and serve as a basis for future research. First, a normal fluid can be a mixture of inviscid and viscous fluid particles. Second, \HLL{while molecular viscosity renders the normal fluid at microscopic scales, its small magnitude contributes little to the large-scale effective viscosity, which includes both molecular and eddy viscosities; therefore, in laminar regimes where eddy viscosity is negligible, the fluid can be approximately treated as inviscid if molecular viscosity is sufficiently low.}

The remainder of this paper is organized as follows: Section 2 reviews our two-fluid model to clarify its significance and position with respect to related methods and theories within the framework of multiscale physics. In Section 3, we discuss the inverse coarse-graining effects of SPH. In particular, we demonstrate the relationship between spatial filtering in LES and smoothing in SPH, showing that they serve as the scale and inverse scale transformations of the two-fluid models based on classical and quantum hydrodynamics, while arguing for the correspondence between the viscosity correction term to conserve the spin angular momentum and SGS model. \HL{In addition, we show that the Condiff viscosity model can serve as an SGS model and incorporate the quantum vortex interactions into the two-fluid model.} Section 4 summarizes the conclusions of this study.

\section{Cryogenic liquid helium-4 in the framework of multiscale physics}
Figure~\ref{fig:Figure-ComparisonWithRelated} presents a schematic of theoretical models describing the dynamics of liquid helium-4, organized by temperature and spatial scale in both physical and discrete spaces. The models are summarized on a single page and embedded in vector format for scalable viewing. Our classical hydrodynamic two-fluid model appears in panels Fig.~\ref{fig:Figure-ComparisonWithRelated}(1)--(3), marked by the red-highlighted label A. An independent version of Fig.~\ref{fig:Figure-ComparisonWithRelated} is available in the Supplementary Material.

\subsection{Microscopic description of liquid helium-4: from Bose--Einstein condensates to quantum hydrodynamics}
We begin with the non-interacting case shown in Fig.~\ref{fig:Figure-ComparisonWithRelated}(b), which outlines basic concepts of bosonic many-body systems. As noted in the Introduction, liquid helium-4 is a bosonic system. The simplest model is the ideal Bose gas, which neglects inter-particle interactions and assumes quantized energy levels with unlimited occupancy by indistinguishable particles. The corresponding statistics follow from quantum statistical mechanics~\cite{Schieve_Horwitz_2009}. The expected ratio of ground-state particles $N_{s}$ to total particles $N_{t}$ at temperature $T$ is given by
\begin{equation}
\frac{N_{s}}{N_{t}} = \left[1 - \frac{T}{T_{B}} \right]^{3/2}, \label{eq:BEC}
\end{equation}
where $T_{B}$ is the critical temperature. This relationship is shown in Fig.~\ref{fig:Figure-ComparisonWithRelated}(15).

Next, we consider Fig.~\ref{fig:Figure-ComparisonWithRelated}(a), which includes interactions between particles. For a symmetric two-body potential $V_{\text{int}}(r - r')$, the Hamiltonian can be expressed via second quantization as:
\begin{equation}
\hat{H} = \int dr \hat{\Psi}^\dagger(r)\left(-\frac{\hbar^2}{2m_{q}} \nabla^2 + V_{\text{ext}}\right)\hat{\Psi}(r) + \frac{1}{2} \int dr \int dr' \hat{\Psi}^\dagger(r)\hat{\Psi}^\dagger(r') V_{\text{int}}(r - r') \hat{\Psi}(r') \hat{\Psi}(r), \label{eq:Hamilton}
\end{equation}
as shown in Fig.~\ref{fig:Figure-ComparisonWithRelated}(14). Here, $\hat{\Psi}$ and $\hat{\Psi}^\dagger$ are field operators for annihilation and creation, respectively; $\hbar$ is the reduced Planck constant; $V_{\text{ext}}$ is an external potential; $V_{\text{int}}$ is the interaction potential; $g$ is the coupling constant; $m_{q}$ is the mass of an atom; and $\delta$ is the Dirac delta function. Second quantization extends creation and annihilation operators to arbitrary positions in space, yielding field operators. A creation field operator at position $r$ is a linear combination of all eigenstates weighted by their respective creation operators and determines their spatial expectation values. A similar formulation applies to annihilation. This framework defines particle number density in space. For further details, see Refs.~\cite{doi:10.1098/rspa.1953.0144, Stoof2009, Schonberg1953}. Understanding the full form of Eqs. (\ref{eq:BEC}) and (\ref{eq:Hamilton}) is not essential here; it suffices to recognize that liquid helium-4 near absolute zero admits such a quantum field-theoretic description in principle.

At finite temperatures, Bose particles are divided into two components: condensed particles in the ground state and non-condensed ones in excited states. Equation (\ref{eq:Hamilton}) outlines the Hamiltonian for the full system including particle--particle interactions. Applying two approximations---(i) the Bogoliubov approximation~\cite{Sankovich2010}, which neglects non-condensed contributions, and (ii) a local interaction model replacing the two-body potential with a Dirac delta function, $V_{\text{int}}(r - r') = g \delta(r - r')$ as shown in Fig.~\ref{fig:Figure-ComparisonWithRelated}(13)---yields the Gross--Pitaevskii (GP) equation:
\begin{equation}
i \hbar \frac{\partial \psi}{\partial t} = \left[ -\frac{\hbar^2 \nabla^2}{2m_{q}} + V_{\text{ext}} - \mu_{\text{GP}} + g|\psi|^2 \right]\psi,
\end{equation}
as represented in Fig.~\ref{fig:Figure-ComparisonWithRelated}(12). The GP equation is a nonlinear Schr${\rm \ddot{o}}$dinger equation governing the time evolution of the condensate wave function~\cite{Gross1961, pitaevskii1961vortex}. While it captures the essential mechanism of helium-4 dynamics, it fails to quantitatively describe strongly interacting systems due to the local interaction assumption in (ii). Thus, alternative approaches are necessary.

In deriving the GP equation, the Bogoliubov approximation neglects contributions from the non-condensed component. While valid near absolute zero, finite temperatures induce excitations, producing a non-condensed fraction with molecular viscosity arising from interatomic interactions---absent in the ground state. Thus, at cryogenic temperatures, two distinct flows exist: a non-viscous superfluid and a viscous normal component. In 1965, I. M. Khalatnikov formalized this two-fluid picture using Newtonian principles: (A) energy flux conservation, (B) the first law of thermodynamics, and (C) Galilean relativity between the two components~\cite{khalatnikov2018introduction}, yielding:
\begin{eqnarray}
\rho_{s} \frac{D \boldsymbol{v}_{s}}{D t} &=& - \frac{\rho_{s}}{\rho} \nabla P + \rho_{s} \sigma \nabla T - \boldsymbol{F}_{sn}, \label{eq:tfmmicroinvisc} \\
\rho_{n} \frac{D \boldsymbol{v}_{n}}{D t} &=& - \frac{\rho_{n}}{\rho} \nabla P - \rho_{s} \sigma \nabla T + \eta \nabla^{2} \boldsymbol{v}_{n} + \boldsymbol{F}_{sn}. \label{eq:tfmmicrovisc}
\end{eqnarray}
Here, $\rho = \rho_{n} + \rho_{s}$ is the total density, $\eta$ the viscosity coefficient, $P$ the pressure, $T$ the temperature, $\sigma$ the entropy density, $\boldsymbol{F}_{sn}$ the mutual friction force, and $\boldsymbol{v}_n$, $\boldsymbol{v}_s$ the velocities of the normal and superfluid components, respectively. These equations follow from substituting (C) and the following conservation laws into (B), then matching both sides of (A):
\begin{eqnarray}
\frac{\partial \rho}{\partial t} + \nabla \cdot \boldsymbol{j} &=& 0, \label{eq:fluxj} \\
\frac{\partial \boldsymbol{j}}{\partial t} + \nabla : (\Pi + \Pi^{\prime}) &=& 0, \label{eq:mmtflux} \\
\frac{\partial \boldsymbol{v}_s}{\partial t} + \nabla (\phi + \frac{1}{2} {\boldsymbol{v}_{s}}^{2} + h^{\prime}) &=& 0, \label{eq:motioneqqm} \\
\frac{\partial S}{\partial t} + \nabla \cdot (\boldsymbol{J}_{S} + \boldsymbol{J}^{\prime}_S) &=& \frac{R}{T} \label{eq:R}.
\end{eqnarray}
This yields $\phi = \mu$ and $\boldsymbol{J}_S = S \boldsymbol{v}_n$, where $\mu$ is the chemical potential and $S$ the entropy. Variables $\Pi^{\prime}$, $h^{\prime}$, and $\boldsymbol{J}^{\prime}_S$ represent dissipation terms; $R$ is the entropy generation rate, and $\boldsymbol{j}$ the momentum density. These relations are illustrated in Figs.~\ref{fig:Figure-ComparisonWithRelated}(8)--(11).

Notably, Eqs. (\ref{eq:fluxj})--(\ref{eq:R}) contain four unknowns---$\Pi ^{\prime}$, $h^{\prime}$, $\boldsymbol{J}^{\prime}_S$, and an additional $\boldsymbol{Q}^{\prime}$ in the energy flux equation---yet only two relations arise from matching (A) and (B) term by term. Thus, these dissipative terms remain undetermined. Typically, they are approximated using classical assumptions: $\Pi^{\prime}$ is modeled via Newtonian viscosity, $\boldsymbol{J}^{\prime}_S$ via Fourier's law ($\boldsymbol{J}^{\prime}_S = -\nabla T$), and $\boldsymbol{Q}^{\prime}$ analogously, despite the model's quantum origin. Furthermore, for the chemical potential $\mu$, the thermodynamic Gibbs--Duhem relation:
\begin{eqnarray}
\nabla \mu = \frac{V}{N}\nabla P - \frac{S}{N} \nabla T, \label{eq:GibbsDuhem}
\end{eqnarray}
is used instead of the expression derived from the GP equation:
\begin{eqnarray}
\mu = g n + V_{\text{ext}} - \frac{\hbar^2}{2m_{q} \sqrt{n}} \nabla^2 \sqrt{n}. \label{eq:chempotGP}
\end{eqnarray}
Here, $S$ is the entropy, $N$ is the number of helium atoms, $n$ is the condensate number density, and $m_{q}$ is the mass of an atom. $S$ and $\sigma$ have the relationship of $\sigma = S/(N m_{q})$. 

It is important to note that the principles underlying the two-fluid model derive from classical Newtonian dynamics. Specifically, energy flux conservation (A), the first law of thermodynamics (B), Galilean relativity (C), and the conservation laws of mass and momentum densities in Eqs. (\ref{eq:fluxj}) and (\ref{eq:mmtflux}) are all standard in classical fluid systems. In contrast, among the resulting relations, $\boldsymbol{J}_S = S \boldsymbol{v}_n$ is unique to cryogenic liquid helium-4, reflecting that only the normal fluid transports entropy. However, this entropy transport picture is also valid for classical systems with inviscid and viscous components.
Crucially, the interaction between the two components is not part of the two-fluid model's derivation. The components were initially considered independent, and mutual friction---later introduced by Gorter and Mellink~\cite{GORTER1949285}---acts oppositely on each component and thus cancels out in total momentum. Hence, it does not violate conservation laws. Even if mutual friction has non-negligible effects, these can be absorbed into the undetermined dissipative terms $\Pi'$, $h'$, and $\boldsymbol{Q}'$, as discussed earlier. In summary, the two-fluid model remains incomplete with respect to both dissipation and inter-component interaction. However, this incompleteness offers flexibility and can be leveraged to extend the model toward multiscale physical descriptions.

\subsection{Macroscopic description of liquid helium-4: the two-fluid model based on classical hydrodynamics}
As noted in the Introduction, our goal was to numerically reproduce macroscopic quantum phenomena---such as torus flow---observed in bulk liquid helium-4. Given their centimeter-scale nature, directly modeling these phenomena using quantum fluid equations, which operate at nanometer to micrometer scales, is computationally impractical. A more feasible approach is to use the Navier--Stokes (NS) equations, appropriate for classical-scale flows, augmented with particle- or quantum-level corrections. From this perspective, liquid helium-4 at cryogenic temperatures can be viewed as a mixture of inviscid and viscous fluids. As temperature decreases, the viscous component diminishes, vanishing entirely at absolute zero to yield a single-phase inviscid flow. We further assume weakly compressible (quasi-incompressible) flow, justified by two points from quantum statistical mechanics: (1) the ratio of superfluid to normal fluid is a fluctuating statistical mean, and (2) even at absolute zero, the condensed component in bulk helium-4 constitutes no more than 13 $\%$ of the total~\cite{PhysRevLett.49.279}. Thus, (1) permits compressibility, and (2) supports the continuum-fluid approximation over a particle-dominated view. We adopted smoothed particle hydrodynamics (SPH) for its ability to accommodate weak compressibility while approximating incompressibility. SPH also introduces useful particle-level corrections. For instance, temperature $T$ is computed as a weighted average over neighboring particles and therefore fluctuates locally, even without explicitly solving its time evolution. These fluctuations in $T$ and density can induce transient temperature gradient forces near fluid interfaces, producing interfacial tension---even when the macroscopic temperature appears uniform. While this effect is negligible under moderate fluctuations, steep gradients can emerge when the two components are in close proximity, making interfacial tension non-negligible under strong external disturbances or localized heating. Additionally, the local interaction force between components can be modeled as proportional to their velocity difference, analogous to fluid--particle interactions in multiphase flow systems~\cite{ROBINSON2014121, HE2018548, Pozorski2024}.

We developed a two-fluid model that (a) comprises classical inviscid and viscous fluid components, (b) includes interfacial tension induced only under exceptional conditions, and (c) introduces a local interaction force proportional to the velocity difference near the interface~\cite{Tsuzuki_2021, doi:10.1063/5.0060605, doi:10.1063/5.0122247, 10.1063/5.0218444}. In essence, this is a Lagrangian two-phase flow model, except for the constrained generation of interfacial tension in (b). The governing equations of our classical hydrodynamic two-fluid model are:
\begin{eqnarray}
\rho_{s} \frac{D \vec{v}_{s}}{D t} &=& - \frac{\rho_{s}}{\rho} \nabla P + \rho_{s} \sigma \nabla T - \vec{F}_{sn}, \label{eq:tfmmacroinvisc} \\
\rho_{n} \frac{D \vec{v}_{n}}{D t} &=& - \frac{\rho_{n}}{\rho} \nabla P - \rho_{s} \sigma \nabla T + \overline{\eta} \nabla^{2} \vec{v}_{n} + \vec{F}_{sn}. \label{eq:tfmmacrovisc}
\end{eqnarray}
Here, $\overline{\eta}$, $\vec{v}_{n}$, $\vec{v}_{s}$, and $\vec{F}_{sn}$ are macroscopic counterparts of the variables $\eta$, $\boldsymbol{v}_{n}$, $\boldsymbol{v}_{s}$, and $\boldsymbol{F}_{sn}$ in Eqs. (\ref{eq:tfmmicroinvisc}) and (\ref{eq:tfmmicrovisc}). Following Condiff's formulation~\cite{doi:10.1063/1.1711295}, the viscosity term can be decomposed into shear, bulk, and rotational contributions with coefficients $\overline{\eta}$, $\overline{\xi}$, and $\overline{\eta}_{r}$, respectively. Redefining $\overline{\eta}$ as the shear coefficient, Eq.~(\ref{eq:tfmmacrovisc}) becomes:
\begin{eqnarray}
\rho_{n} \frac{D \vec{v}_{n}}{D t} &=& - \frac{\rho_{n}}{\rho} \nabla P - \rho_{s} \sigma \nabla T + (\overline{\eta}  +\overline{\eta_{r}} ) \nabla ^{2} \vec{v}_n+ (\frac{\overline{\eta}}{3}  +\overline{\xi} - \overline{\eta_{r} } )\nabla \nabla \cdot \vec{v}_n  + 2 \overline{\eta_{r} } \nabla \times \vec{\omega}_{0} + \vec{F}_{sn}. \label{eq:tfmmacrocondiff}
\end{eqnarray}
Equations~(\ref{eq:tfmmacroinvisc})--(\ref{eq:tfmmacrocondiff}) are depicted in Figs.~\ref{fig:Figure-ComparisonWithRelated}(1)--(3) beside the red-highlighted label A. We refer to the components in Eqs. (\ref{eq:tfmmacroinvisc})--(\ref{eq:tfmmacrovisc}) as ``inviscid fluid'' and ``viscous fluid,'' respectively, to distinguish them from the ``superfluid'' and ``normal fluid'' of the quantum model in Eqs. (\ref{eq:tfmmicroinvisc}) and (\ref{eq:tfmmicrovisc}). These labels were revised from our earlier naming~\cite{Tsuzuki_2021, doi:10.1063/5.0060605, doi:10.1063/5.0122247} to clarify differences in spatial scale. Importantly, Eqs. (\ref{eq:tfmmacroinvisc}) and (\ref{eq:tfmmacrovisc}) maintain the same structural form as their quantum counterparts, Eqs. (\ref{eq:tfmmicroinvisc}) and (\ref{eq:tfmmicrovisc}). This consistency arises because their derivation relies only on Newtonian principles: (A) energy flux conservation, (B) the first law of thermodynamics, (C) Galilean invariance, and the conservation laws for mass, momentum, and entropy---as shown in Eqs. (\ref{eq:fluxj})--(\ref{eq:R}). Furthermore, as with the mutual friction terms in the quantum model, the interaction forces in the classical model act in opposite directions and cancel, ensuring consistency across scales despite differences in parameter definitions. Thus, Eqs. (\ref{eq:tfmmacroinvisc}) and (\ref{eq:tfmmacrovisc}) are effectively obtained by relabeling ``superfluid'' as ``inviscid fluid'' and ``normal fluid'' as ``viscous fluid,'' and redefining variables such as $\eta$, $\boldsymbol{v}_{n}$, $\boldsymbol{v}_{s}$, and $\boldsymbol{F}_{sn}$ at macroscopic scales.

Equation (\ref{eq:tfmmacrocondiff}) re-derives the viscosity term, following Condiff's reformulation of the Navier--Stokes equation for polar fluids~\cite{doi:10.1063/1.1711295}, in which viscous stresses are decomposed into shear ($\overline{\eta}$), bulk ($\overline{\xi}$), and rotational viscosity ($\overline{\eta}_r$). The fifth term on the right-hand side includes the parameter $\vec{\omega}_0$, representing internal rotational degrees of freedom of a constituent particle---interpreted as spin angular velocity if the particle is modeled as a rigid sphere. By treating $\vec{\omega}_0$ as constant, the model preserves rotational velocity around the particle's axis. Although originally defined for molecules, $\vec{\omega}_0$ has since been adopted in coarse-grained fluid models and interpreted as a virtual fluid particle parameter~\cite{MULLER2015301}. Conservation of spin angular momentum via Eq. (\ref{eq:tfmmacrocondiff}) may bridge classical and quantum hydrodynamics by enabling angular momentum quantization per particle. In this view, enforcing spin angular velocity quantization in the local flow field governing vortex dynamics could approximate quantum hydrodynamic behavior. While this remains a hypothesis, our recent simulations---applying an SPH-discretized, spin-angular-momentum-conserving two-fluid model---successfully reproduced velocity profiles in vortex lattice and counterflow phenomena previously attributed solely to quantum hydrodynamics, lending support to this idea. For further discussion of Eq. (\ref{eq:tfmmacrocondiff})'s derivation and physical meaning, see Sec. II.A and Fig. 9 of Ref.~\cite{10.1063/5.0218444}. Its application to 2D helium-4 flow appears in Refs.~\cite{Tsuzuki_2021, doi:10.1063/5.0060605, doi:10.1063/5.0122247}, and the original SPH discretization of Eq. (\ref{eq:tfmmacrocondiff}) is found in Ref.~\cite{MULLER2015301}.

We now compare the physical significance of density $\rho$ and viscosity $\eta$ in the two-fluid models based on classical hydrodynamics [Eqs. (\ref{eq:tfmmacroinvisc}), (\ref{eq:tfmmacrovisc})] and quantum hydrodynamics [Eqs. (\ref{eq:tfmmicroinvisc}), (\ref{eq:tfmmicrovisc})]. Since density is defined as mass per unit volume, its meaning is consistent across both models. However, $\rho$ is only well-defined once field operators and local averaging are introduced, implying it applies only at spatial scales larger than those governed by the GP equation. In contrast, the interpretation of viscosity differs fundamentally. In the classical hydrodynamic two-fluid model, viscosity coefficients---$\overline{\eta}$ in Eq. (\ref{eq:tfmmacroinvisc}), or $\overline{\eta}$, $\overline{\xi}$, and $\overline{\eta}_r$ in Eq. (\ref{eq:tfmmacrocondiff})---are continuum mechanical parameters defined as the ratio of stress to velocity gradient. Conversely, in the quantum hydrodynamic two-fluid model, $\eta$ is a microscopic transport coefficient relevant in transitional or free molecular flow regimes described by the Boltzmann equation, rather than the Navier--Stokes regime. According to Maxwell's theory~\cite{Maxwell01011860, Brush1962, maxwell1986maxwell}, the viscosity of an ideal gas can be expressed as $\eta = \rho \lambda \sqrt{\overline{v^2}}/3$, where $\lambda$ is the mean free path and $\sqrt{\overline{v^2}}$ is the root mean square molecular velocity. Thus, $\eta$ has different definitions at microscopic and macroscopic levels. Nevertheless, as molecular viscosity is an intrinsic material property, the numerical values in Eqs. (\ref{eq:tfmmicrovisc}) and (\ref{eq:tfmmacrovisc}) may be expected to match, i.e., $\eta = \overline{\eta}$. From a kinetic standpoint, eddy viscosity offers a macroscopic measure of apparent viscosity in high Reynolds number flows. Understanding the onset of turbulence could provide further insights into large-scale helium-4 dynamics. As a first step, the following section explores the relationship between these two-fluid models from a multiscale physics perspective.

\section{Discussion}
There is a clear distinction between vortex dynamics in the classical and quantum hydrodynamic regimes; in the former, vortices dissipate, whereas in the latter, vortices do not dissipate and exist in a stable state because the circulation is quantized. The large vortices observed on the spatial scale of classical hydrodynamic cascade into smaller vortices. At this time, energy passes from the larger vortex to the smaller vortex. Eventually, when the vortex size reaches the Kolmogorov microscale~\cite{KATOPODES2019566, STEINBERG2021100900}, it dissipates and disappears owing to molecular viscosity. While larger vortex sizes tend to exhibit characteristics specific to individual problems, as the vortex size decreases, the vortex becomes more universal in nature, independent of dissipation and external forces. The regime in which such vortices are observed is called a universal subrange; in particular, the inertia-dominated regime is called an inertial subrange zone~\cite{KATOPODES2019566, DOVIAK1993424}. However, small vortices sometimes merge to form large vortices known as inverted cascades. In other words, from a macroscopic perspective, vortex dynamics are a statistical phenomenon. In general, in a uniform and isotropic classical fluid, the relationship between the vortex energy spectrum $E$ and wavenumber $k$ is $E(k) = k^{\frac{-5}{3}}$ in the inertial regime (Kolmogorov's $-5/3$ law)~\cite{kolmogorov1941local}. In quantum hydrodynamics, vortices do not dissipate because the circulation is quantized as previously mentioned. However, the energy transfer mechanism is similar to that of classical fluid systems, especially in inertial subrange zones, where the Kolmogorov's -5/3 law has been verified in recent studies. In other words, the property observed in classical fluid dynamics, cascade splitting from large vortices to small vortices and the accompanying energy transfer, is also a common property in quantum fluid dynamics under certain conditions. In summary, the flow structures in the classical and quantum hydrodynamic regimes exhibit certain common properties.

In classical hydrodynamics, vortices dissipate below the Kolmogorov scale due to molecular viscosity. However, this mechanism fails in quantum fluids like liquid helium-4 at cryogenic temperatures, where molecular viscosity is negligible---even at sub-Kolmogorov scales. Thus, the connection between classical turbulent structures and quantum turbulence, in which quantized vortices are stable and non-dissipative, remains unclear. Specifically, how eddy energy transfer and vortex splitting or merging evolve across the transition from classical to quantum scales is not yet well understood. The space--time diagram in Fig.~\ref{fig:Figure-ComparisonWithRelated} offers key insights into the multiscale physics of helium-4. Across spatial scales, the governing dynamics can be consistently described by a two-fluid model. The formulations in Eqs. (\ref{eq:tfmmacroinvisc})--(\ref{eq:tfmmacrovisc}) and (\ref{eq:tfmmicroinvisc})--(\ref{eq:tfmmicrovisc}) differ primarily in scale, which is not explicitly encoded in the equations. Equation (\ref{eq:tfmmacrocondiff}), however, is essential for capturing quasi-quantum effects within the classical fluid regime. Here, we examine the relationship between the two-fluid model based on quantum hydrodynamics [Eqs. (\ref{eq:tfmmicroinvisc}), (\ref{eq:tfmmicrovisc})] and that based on classical hydrodynamics [Eqs. (\ref{eq:tfmmacroinvisc}), (\ref{eq:tfmmacrovisc})]. We show that these models are connected via scale transformations: classical-to-quantum through filtering, as in large eddy simulation (LES), and quantum-to-classical through inverse transformations, as realized via SPH. Notably, the spin angular momentum conservation term---originally introduced as a quantum correction---can be formally interpreted as a subgrid-scale (SGS) model resulting from this transformation. 
\HLLL{Moreover, solving the classical hydrodynamic two-fluid model with SPH appears to reproduce microscopic-scale fluctuations.}
The degree of fluctuation scales with the kernel radius. \HLLL{This may result from kernel smoothing truncation errors that resemble such fluctuations. However, this resemblance lacks first-principle justification and should be viewed as a speculative analogy rather than a physically grounded effect.}
We also propose that the Condiff viscosity model serves as an SGS framework that embeds quantum vortex interactions into the two-fluid formulation. This multiscale perspective offers new insights into the microscopic structure of cryogenic helium-4. First, the ``normal fluid'' can be interpreted as a mixture of inviscid and viscous fluid particles. Second, \HLL{while molecular viscosity renders the normal fluid at microscopic scales, its small magnitude contributes little to the large-scale effective viscosity, which includes both molecular and eddy viscosities; therefore, in laminar regimes where eddy viscosity is negligible, the normal fluid may be effectively treated as inviscid at large scales if molecular viscosity is sufficiently small}.

\subsection{Correspondence between spatial filtering in LES and kernel approximation in SPH}
In LES, the physical quantity $f$ is decomposed into the resolved (filtered) component $\overline{f}$ and the fluctuation component ${f}_{\epsilon}$:
\begin{eqnarray}
f = \overline{f} + {f}_{\epsilon}, \label{eq:LESconcept} 
\end{eqnarray}
where the resolved component $\overline{f}$ is obtained via spatial filtering as follows~\cite{mcdonough2007introductory}:
\begin{eqnarray}
\overline{f}(\vec{r},t)&\equiv&\int f(\acute{\vec{r}}, t)G(\vec{r}-\acute{\vec{r}}, \Delta) d\acute{\vec{r}}. \label{eq:LESfiltering}
\end{eqnarray}
Here, we omit the position $\vec{r}$ and time $t$ in Eq.~(\ref{eq:LESfiltering}) for easy explanation. 
A simple calculation using Eq.~(\ref{eq:LESfiltering}) reveals the following properties with respect to $\overline{f}$ and $f_{\epsilon}$~\cite{Deardorff_1970, LEONARD1975237}:
\begin{eqnarray}
\overline{\frac{\partial f}{\partial x_{i}}} ~~=~~ \frac{\partial \overline{f}}{\partial x_{i}}, 
~~~~\overline{\overline{f}} ~~\ne~~ \overline{f}, 
~~~~\overline{f_{\epsilon}}~~\ne~~0.
\label{eq:LESrelation}
\end{eqnarray}
The properties in Eq.~(\ref{eq:LESrelation}) epitomize the filtering operations in the LES, which differs from the ensemble average used in the Reynolds--averaged Navier--Stokes (RANS) equation~\cite{doi:10.1098/rsta.1895.0004}. 
In addition, $G(\vec{r}-\acute{\vec{r}}, \Delta)$ in Eq.~(\ref{eq:LESfiltering}) represents a filter function with width $\Delta$ that satisfies the normalization condition $\int G(\vec{r}-\acute{\vec{r}}, \Delta) d\acute{\vec{r}} = 1$. An example of $G$ is the Gaussian, which is expressed as 
\begin{eqnarray}
G(\vec{r} -\acute{\vec{r}}, \Delta)&:=& \frac{C}{{\Delta}^{m}}{\rm e}^{-(\vec{r} - \acute{\vec{r}})^2/{\Delta}^{2}}, \label{eq:gauss}
\end{eqnarray}
where $C$ represents a normalization factor, and $m$ indicates the dimension ($m$ = 1, 2, or 3).

In SPH, a physical quantity $f$ expressed in integral form using the Dirac delta function can be approximated by replacing the delta function with a kernel function $W$, which is a smoothing function with a width of $h$:
\begin{eqnarray}
f (\vec{r}, t) &=& \int f (\acute{\vec{r}}, t) \delta(\vec{r} - \acute{\vec{r}})d\acute{\vec{r}}~~\simeq~~\int f (\acute{\vec{r}}, t) W(\vec{r} - \acute{\vec{r}}, h)d\acute{\vec{r}}. \label{eq:SPHapprox}
\end{eqnarray}
The kernel function $W$ must satisfy the normalization condition $\int W (\vec{r} - \acute{\vec{r}}, h) d\vec{r} = 1$. Additionally, $W$ must be an even function as $W(\vec{r}) = W(-\vec{r})$. There are several types of kernel functions that satisfy these conditions~\cite{desbrun1996smoothed, muller2003particle}. 
\HLL{The Gaussian function is a typical example of a kernel function. Polynomial kernels are often preferred because they possess compact support, vanishing at integer multiples of a parameter $h$. In contrast, the Gaussian kernel lacks this property, extending over an infinite domain and treated as approximately normalized. While compact support is essential in incompressible SPH simulations---favoring polynomial kernels---the Gaussian kernel is commonly used for compressible fluids due to its analytical form and broader effective range. In this study, we focus on the Gaussian kernel, as its analytical nature facilitates theoretical modeling, and liquid helium, being weakly compressible due to density variations, permits relaxing the compact support condition.}

We can estimate the approximation error of replacing the second equation with the third equation in Eq.~(\ref{eq:SPHapprox}) as follows. For explanation, let us express the second equation as $f$ because it describes the true value, and the third equation as $\overline{f}$ because it represents a filtered value of $f$ by the kernel function $W$, as follows:
\begin{eqnarray}
f &\coloneqq& \int f (\acute{\vec{r}}) \delta(\vec{r} - \acute{\vec{r}})d\acute{\vec{r}},  \label{eq:deftruef} \\ 
\overline{f} &\coloneqq& \int f (\acute{\vec{r}}) W(\vec{r} - \acute{\vec{r}}, h) d\acute{\vec{r}}. \label{eq:defoverf}
\end{eqnarray}
Here, we omit the position and time on the left-hand sides of Eqs.~(\ref{eq:deftruef}) and (\ref{eq:defoverf}).
In the one-dimensional case, the Taylor expansion of $\overline{f}$ yields the following~\cite{STRANEX2011392}:
\begin{eqnarray}
\overline{f} =  M_{0} f - h M_{1} f^{(1)} + h^{2} \frac{M_{2}}{2} f^{(2)} -\cdots,
\end{eqnarray}
where $f^{(k)}$ is the $k$th derivative of $f$, and $M_{k}$ represents the $k$th moment of kernel $W$ expressed as
\begin{eqnarray}
M_{k} = \int {\rm r}^{k} W({\rm r}) d {\rm r}, \label{eq:SPHmomentfunc}
\end{eqnarray}
where ${\rm r} = \frac{{\rm x}-\acute{{\rm x}}}{h}$ in the $ x $ direction in the one-dimensional problem. 
$M_{0}$ is always 1 because of the normalization condition $\int W({\rm r}) d{\rm r}=1$. Therefore, we obtain the following relationship:
\begin{eqnarray}
\overline{f} =  f - h M_{1} f^{(1)} + h^{2} \frac{M_{2}}{2} f^{(2)} -\cdots, \label{eq:SPHerror1dim}
\end{eqnarray}
where the odd terms always vanish because $M_{k}$ becomes zero when $k$ is odd.
Equation~(\ref{eq:SPHerror1dim}) can be extended to a general form in multiple dimensions as follows~\cite{Gabbasov_2017, SIGALOTTI201950}:
\begin{eqnarray}
\overline{f} &=&  f + \displaystyle \sum _{l=1}^{\infty }\displaystyle \frac{1}{l!}{{\rm{\nabla} }}^{(l)}f({\vec{r}})\,:::\cdots :{\displaystyle \int } {({\acute{\vec{r}}}-{\vec{r}})}^{l} W({\vec{r}}-{{\acute{\vec{r}}}},h){d}^{m}{\acute{\vec{r}}} \label{eq:SPHerrormuldim} \\ 
&=& f + f_{\epsilon}^{{\rm SPH}}. \label{eq:SPHerrortheref}
\end{eqnarray}
Here, the symbol ``$:::\cdots :$'' represents the $l$th order inner product, and $m$ indicates the dimension ($m$ = 1, 2, or 3).  
We have denoted the second term on the right-hand side of Eq.~(\ref{eq:SPHerrormuldim}) as $f_{\epsilon}^{{\rm SPH}}$ in Eq.~(\ref{eq:SPHerrortheref}).
Notably, the arguments in Eqs. (\ref{eq:defoverf})--(\ref{eq:SPHerrortheref}) remain valid when the kernel function $W$ is replaced with $G$. This is because only the normalization condition of the $n$th-order differentiability is required for $W$ and $G$ as filter or smoothing functions, and imposing this condition does not restrict the selectivity of $W$ and $G$. 

We describe $f_{\epsilon}$ in Eq.~(\ref{eq:LESconcept}) as $f_{\epsilon}^{{\rm LES}}$. The expansion of Eq.~(\ref{eq:LESfiltering}) in the Taylor series, as in Eq.~(\ref{eq:defoverf}), leads to Eq.~(\ref{eq:SPHerrormuldim}) by replacing $W$ with $G$. Here, $W = G$ holds true when the Gaussian is designated to both of them; in this case, $f_{\epsilon}^{\rm SPH} = - f_{\epsilon}^{\rm LES}$ is established. In summary, the following relationships were obtained:
\begin{eqnarray}
f &=& \overline{f} + {f}_{\epsilon}^{{\rm LES}} \label{eq:redefoneles} \\
\overline{f} &=& f + f_{\epsilon}^{{\rm SPH}} 
\label{eq:smrlessph}
\end{eqnarray}
A previous study has indicated that SPH can be reinterpreted as a filtering of the LES~~\cite{10.1063/1.4978274}. This study admits the mathematical equivalence between the smoothing of the SPH and filtering of the LES, and we present a new physical interpretation of both in multiscale physics of the cryogenic liquid helium-4: First, Eq.~(\ref{eq:redefoneles}) represents the transformation equation in the LES from microscopic to macroscopic scale. The profile is a true value $f$ that follows the governing equations at a small scale. Filtering eliminates the fluctuation $f_{\epsilon}^{\rm LES}$ at the small scale by smoothing $f$, resulting in $\overline{f}$ and a system of fluid equations that $\overline{f}$  follows at the large scale. Conversely, Eq.~(\ref{eq:smrlessph}) represents the transformation equation for the SPH operation between the macroscopic value $\overline{f}$ and the microscopic value $f$; in this case, the governing equations are the hydrodynamic equations that $\overline{f}$ follows on a large scale. However, in SPH computation, the physical quantity possessed by each particle is $f$, which is a small-scale quantity. $\overline{f}$ is calculated on the large scale using the weight calculation on the right-hand side of Eq.~(\ref{eq:defoverf}) as a smoothing operation. However, $\overline{f}$ always includes the fluctuation $f_{\epsilon}^{\rm SPH}$ caused by the smoothing kernel approximation error, as shown in Eq.~(\ref{eq:smrlessph}). Consequently, the microscopic fluctuations $f_{\epsilon}^{\rm SPH}$ can be generated in the governing equations at the macroscopic scale. Here, $f_{\epsilon}^{\rm SPH}$ is proportional to the sum of the even powers of the smoothing width $h$, as shown on the right-hand side of Eq.~(\ref{eq:SPHerror1dim}). Therefore, as width $h$ increases, so does $f_{\epsilon}^{\rm SPH}$. Consequently, $\overline{f}$ contains an amplified version of $f_{\epsilon}^{\rm SPH}$ compared with the original value, thus reproducing the microscopic behavior even at different large scales. Briefly, the SPH form serves as a magnifying glass for microscopic phenomena. Therefore, macroscopic quantum phenomena were replicated in our simulations, even by solving a system of equations on the hydrodynamic scale.

The SPH form has two errors: the smoothing kernel approximation and particle approximation errors. The smoothing kernel approximation error is inherent in the SPH form and exists as long as the width $h$ is within the range of finite values. This error is inherent in the theory of approximating the Dirac delta function with a finite-width distribution. In contrast, the particle approximation error is a discretization error that occurs when the kernel function is reproduced on a computer. If the regularity and $h$-connectivity conditions are satisfied, the discretization error can be minimized by increasing the resolution~\cite{Imoto2019, Imoto2020}. Thus, the two error types exhibit different characteristics. We have omitted the particle approximation error from the discussion here because it is a computational error that can be minimized by performing high-resolution simulations on supercomputers if the appropriate discrete SPH forms are employed. Nevertheless, this argument can be extended to include the particle approximation. The relationship between kernel approximation errors, particle approximation errors, and true values in SPH is reported in Ref.~\cite{AMICARELLI2011279}. Let $f$ be the true value, $f^{s}$ the kernel approximation value of $f$, and $f^{p}$ the particle approximation value of $f^{s}$. Additionally, let $f_{\epsilon}^{s}$ be the kernel approximation error, $f_{\epsilon}^{p}$ the particle approximation error, and $f_{\epsilon}$ the entire truncation error. We have the relationship $f_{\epsilon} = $ $(f^{p} - f) = $ $(f^{s} - f) + (f^{p} - f^{s}) = $ $(f_{\epsilon}^{s} + f_{\epsilon}^{p})$. Accordingly, $f_{\epsilon}^{\rm SPH}$ in Eq.~(\ref{eq:smrlessph}) can be reinterpreted to include the particle approximation error. In this case, $f_{\epsilon}^{SPH}$ does not necessarily correspond to $- f_{\epsilon}^{\rm LES}$ because it is established only when kernel approximation errors are considered. In addition, numerical instability produces unpredictable errors in actual calculations~\cite{BALSARA1995357, SWEGLE1995123}. The equality between $f_{\epsilon}^{\rm SPH}$ and $- f_{\epsilon}^{\rm LES}$ is a formal relationship that holds when the particle approximation and numerical errors are ignored. In summary, we showed the equivalence between the spatial filtering of the LES and the kernel approximation of SPH; both are formally related as scale and inverse scale transformations in multiscale physics.

\HLL{Based on the above considerations, we have shown that the truncation error $f_{\epsilon}^{\rm SPH}$ formally corresponds to the subscale fluctuation term $f_{\epsilon}^{\rm LES}$. However, for $f_{\epsilon}^{\rm SPH}$ to effectively substitute for the physical behavior represented by $f_{\epsilon}^{\rm LES}$, several conditions must be satisfied.
First, the value of $f_{\epsilon}^{\rm SPH}$ must be { \it probabilistic}. This requirement arises because $f_{\epsilon}^{\rm SPH}$, while inherently uncertain as a remainder term from kernel approximation, is deterministic in nature. In contrast, $f_{\epsilon}^{\rm LES}$ reflects microscopic physical fluctuations (e.g., Brownian motion) and is intrinsically probabilistic. Since the kernel function itself is deterministic, in order for $f_{\epsilon}^{\rm SPH}$ to acquire a probabilistic nature through its indeterminacy, the physical quantity $f$ and its derivatives must be probabilistically influenced by the statistical variability in particle distribution. \HLL{In particle-based methods such as SPH and moving particle semi-implicit (MPS), a few percent statistical variance in density is typically observed, and physical quantities are affected by this variability, thereby acquiring a probabilistic character. Thus, this condition can be satisfied in such frameworks. Conversely, fluid solvers that do not incorporate statistical fluctuations in physical quantities cannot fulfill this condition.}
Second, $f_{\epsilon}^{\rm SPH}$ and $f_{\epsilon}^{\rm LES}$ must share the same statistical properties. According to Eq.~(\ref{eq:LESrelation}), we require that $\overline{f_{\epsilon}^{\rm SPH}} \ne 0$.
In summary, if the following three conditions are met:
(i) the physical quantity $f$ and its derivatives become probabilistic due to the statistical variation in particle distribution, and this mimics the true physical behavior of $f$,
(ii) the resulting $f_{\epsilon}^{\rm SPH}$ satisfies the same statistical property as $f_{\epsilon}^{\rm LES}$, namely $\overline{f_{\epsilon}^{\rm SPH}} \ne 0$, and
(iii) the physical characteristics represented by $f_{\epsilon}^{\rm SPH}$ and $f_{\epsilon}^{\rm LES}$ are similar in the target problem, then $f_{\epsilon}^{\rm SPH}$ can serve as a high-probability substitute for $f_{\epsilon}^{\rm LES}$ within that problem domain.}

\HLL{It is generally difficult to prove the universal validity of conditions (i)--(iii). Nevertheless, specific examples can be considered. Since $f_{\epsilon}^{\rm LES}$ represents the fine-scale fluctuation of a physical quantity $f$, it is often reasonable to assume that $f$ exhibits white noise characteristics, particularly in isolated or isotropic systems. Furthermore, consider the convolution of white noise $\epsilon(x)$ with a finite-width distribution function $\varphi(x)$: $\overline{\epsilon(x)} := \int_{-\infty}^{\infty} \epsilon(s) \cdot \varphi(x - s) ds.$
Here, $\overline{\epsilon(x)}$ represents a weighted average of white noise, which remains random and possesses nonzero variance. Therefore, in general, $\overline{\epsilon(x)} \ne 0$. Hence, condition (ii) is naturally satisfied in the case of white noise. Thus, the proposition to be demonstrated becomes: If the physical quantity $f$ exhibits white noise characteristics, does the truncation error $f_{\epsilon}^{\rm SPH}$ arising from kernel approximation of $f$ also exhibit white noise characteristics ?}

\HLL{For simplicity, we consider a one-dimensional problem and truncate the Taylor expansion in Eq.~(\ref{eq:SPHerrormuldim}) at second order. Let $f$ be a stochastic function, and assume that its third derivative is spatially uncorrelated Gaussian noise from the perspective of the maximum entropy:
\begin{eqnarray}
\frac{d^3 f}{dx^3}(x) := \eta(x), \quad \eta(x) \sim \mathcal{N}(0, \sigma_{p}^2),
\end{eqnarray}
where $\sigma_{p}^2$ denotes the variance of the white noise. Then, the remainder term $f_{\varepsilon}^{(2)}(x)$ of the third-order Taylor expansion takes the following form:
\begin{eqnarray}
f_{\varepsilon}^{(2)}(x) = \frac{1}{6} \int r^3 \, \eta(x + \theta r) \, W(r, h) \, dr.
\end{eqnarray}
Here, the kernel function is taken to be Gaussian: $W(r, h) = \frac{1}{\sqrt{2\pi}h} \exp\left(-\frac{r^2}{2h^2}\right)$, and the integration point $\theta$ is chosen from the interval $(0, 1)$. For $f_{\varepsilon}^{(2)}(x)$ to be considered white noise, it must satisfy $\mathbb{E}[f_{\varepsilon}^{(2)}(x)] = 0$ and exhibit spatial autocorrelation of the Dirac delta form: $\mathbb{E}[f_{\varepsilon}^{(2)}(x) f_{\varepsilon}^{(2)}(x')] = \sigma_{p}^2 \delta(x - x')$.}

\begin{figure*}[t]
\vspace{-5.0cm}
\hspace{+3.2cm}
\centerline{\includegraphics[width=1.3\textwidth, clip, bb= 0 0 1600 800 ]{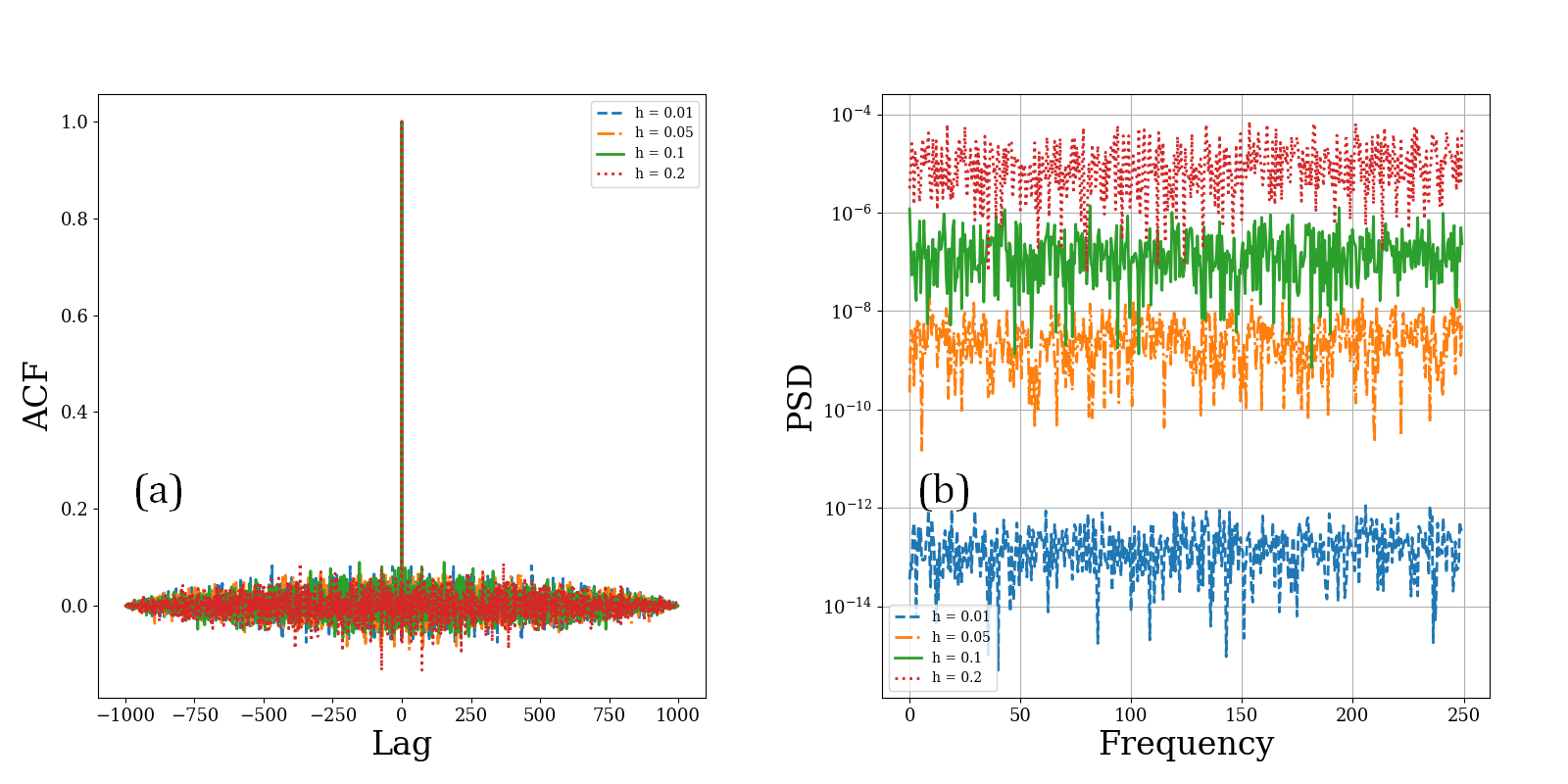}}
\caption{\HLL{(a) Autocorrelation function (ACF) and (b) power spectral density (PSD) of the truncation error $f_{\varepsilon}^{(2)}(x)$ from SPH kernel approximation, estimated via Monte Carlo simulation for various kernel widths $h \in \{0.01, 0.05, 0.1, 0.2\}$. The ACF exhibits a delta-like profile centered at zero, while the PSD remains nearly flat across frequencies, indicating white noise characteristics.}}
\label{fig:Figure-SpectralAnalysis}
\end{figure*}

\HLL{To verify this, we evaluated the Monte Carlo estimator of the above integral for various kernel widths $h$. Specifically, we set the number of spatial subdivisions on the interval $[-1, 1]$ to $M = 1000$, and the number of samples per division to $N = 2000$, i.e., $(M, N) = (1000, 2000)$. Using Monte Carlo simulations, we generated samples of $f_{\varepsilon}^{(2)}(x)$ for several kernel widths $h \in \{0.01, 0.05, 0.1, 0.2\}$, and computed the corresponding autocorrelation functions (ACFs) and power spectral densities (PSDs). The results are shown in Figs.~\ref{fig:Figure-SpectralAnalysis}(a) and (b), with (b) plotted on a semi-logarithmic scale.
An ACF that vanishes everywhere except at the center indicates a delta function-like behavior, while low frequency dependence in the PSD implies strong white-noise characteristics. The results in (a) and (b) collectively demonstrate that when the physical quantity $f$ exhibits white noise behavior, the truncation error $f_{\epsilon}^{\rm SPH}$ resulting from its kernel approximation also displays white noise characteristics.
These findings suggest that, under certain conditions, the numerical errors arising in computational models may serve as reasonable proxies for fine-scale physical fluctuations such as Brownian motion.
In particular, as shown in Figs.~\ref{fig:Figure-SpectralAnalysis}(a) and (b), white noise characteristics were observed in all cases, irrespective of the kernel radius $h$. Since $f_{\epsilon}^{\rm SPH}$ is expanded in even powers of $h$, as indicated by Eq. (\ref{eq:SPHerror1dim}), larger values of $h$ result in small-scale variance being described by variance at larger scales. However, $h$ must satisfy a regularity condition~\cite{Imoto2020}---namely, the number of particles within the support radius must exceed a certain threshold. If not, the Taylor expansion loses convergence, and the expansion of physical quantities such as $f$ becomes unreliable. In short, to amplify and capture microscopic fluctuations, a larger $h$ is required, but this in turn demands higher particle density to maintain regularity, increasing the computational resolution. In contrast, as resolution increases, the particle approximation error $f_p$ decreases and eventually becomes negligible, and the SPH error is then primarily governed by the truncation error $f_{\epsilon}^{\rm SPH}$. Further study is needed to clarify how the regularity condition, the optimal magnification of $f_{\epsilon}^{\rm SPH}$ via $h$, and particle size are interrelated.}

\HLLL{The fluctuations observed in SPH simulations primarily originate from numerical artifacts such as truncation errors in kernel approximations, summation inconsistencies, and particle disorder. While these artificial fluctuations may, under certain conditions, resemble physical microscopic behavior---such as exhibiting white-noise characteristics---we do not suggest that they intrinsically follow quantum mechanical laws or reflect fundamental physical processes. Rather, such resemblance is considered coincidental and context-dependent. Establishing a rigorous causal link between numerical fluctuations in SPH and genuine quantum features remains an open and important question for future investigation.}

\HLL{In addition, we now discuss the connectivity between classical and quantum fluids. In recent work, the author showed that within the SPH framework, the quantum fluid equation and the incompressible inviscid fluid equation (i.e., the Euler equation) become equivalent when the spatial gradient of density is sufficiently moderate and the SPH fluid particles do not exceed Landau's critical velocity~\cite{doi:10.1063/5.0122247}. More specifically, under these conditions, the SPH discretized form of the quantum fluid equation---derived from Eq. (\ref{eq:motioneqqm}) by expressing the scalar function $\phi$ as the quantum mechanical chemical potential $\mu$ in Eq. (\ref{eq:chempotGP}), while neglecting both the dissipation term $h'$ and the quantum pressure term---becomes equivalent to the inviscid component of the two-fluid model in Eq. (\ref{eq:tfmmicroinvisc}) , with the mutual friction term $\boldsymbol{F}_{sn}$ omitted. This equivalence arises from the identity:}
\begin{eqnarray}
-\sum_{j=1}^{N_p} \biggr(\frac{P_j}{\rho} - \sigma T_j\biggl)\frac{m_j}{\rho_j} \nabla W_{ij} = -\sum_{j=1}^{N_p} \biggr(\frac{gn_j + \HLLL{V_{\text{ext}}^{j}}}{m_q}\biggl)\frac{m_j}{\rho_j} \nabla W_{ij},
\end{eqnarray}
which holds under the aforementioned conditions. From this identity, a thermodynamic relation is obtained for each particle:
\begin{eqnarray}
\therefore ST_{j} - VP_{j}+ N\mu_{j} = 0.
\end{eqnarray}
Here, $N_p$ denotes the number of virtual fluid particles in SPH, $N$ is the number of helium atoms, and $g$ is the coupling constant. $\rho$ represents the mass density. The variables $\rho_j$, $P_j$, $T_j$, $m_j$, $n_j$, and \HLLL{$V_{\text{ext}}^j$} denote, respectively, the density, pressure, temperature, mass, condensate number density, and external potential of the $j$th virtual fluid particle. Importantly, the above equation is a special case of the thermodynamic Euler equation $S T_j - V P_j + N \mu_j = U_j$ with $U_j = 0$, \HLLL{where $U_j$ is the internal energy of the $j$th particle}. In other words, when the density gradient is sufficiently small and the SPH particles \HLLL{remain below} Landau's critical velocity---\HLLL{so that} the quantum pressure and mutual friction terms \HLLL{can be} neglected---the internal energy of each fluid particle must vanish ($U_j = 0$) for the SPH-discretized quantum fluid equation to be equivalent to the incompressible Euler equation. Under these assumptions, microscopic-scale \HLLL{dynamics} behave identically in both the quantum and classical (Euler) formulations, \HLLL{lending support to the notion} that \HLLL{the microscopic dynamics} can propagate to larger scales \HLLL{within the framework of classical hydrodynamics}. 

\subsection{Subgrid-scale (SGS) model for the multi-scale cryogenic liquid helium-4} \label{sec:closure}
Suppose that $\rho$, $\rho_{s}$, $\rho_{n}$, $\sigma$, and $\eta$ in Eqs. (\ref{eq:tfmmicroinvisc}) and (\ref{eq:tfmmicrovisc}) are constant parameters. By decomposing $\boldsymbol{v}_{s}$, $\boldsymbol{v}_{n}$, $P$, $T$, and $\boldsymbol{F}_{sn}$ into the filtered and fluctuation components as $\boldsymbol{v}_{s} = $ $\overline{\boldsymbol{v}}_{s} + \boldsymbol{v}_{\epsilon}^{(s)}$, $\boldsymbol{v}_{n} = $ $\overline{\boldsymbol{v}}_{n} + \boldsymbol{v}_{\epsilon}^{(n)}$, $P = $ $\overline{P} + P_{\epsilon}$, $T = $ $\overline{T} + T_{\epsilon}$, and $\boldsymbol{F}_{sn} = $ $\overline{\boldsymbol{F}}_{sn} + \boldsymbol{F}_{\epsilon}$, and by filtering Eqs. (\ref{eq:tfmmicroinvisc}) and (\ref{eq:tfmmicrovisc}) using Eq.~(\ref{eq:LESfiltering}), we obtain a system of equations for the two-fluid model at large scales, as follows:
\begin{eqnarray}
{\rho }_{s}\displaystyle \frac{{D}{{\boldsymbol{\overline{v}}}}_{s}}{{D}t}&=& -\displaystyle \frac{{\rho }_{s}}{\rho }{\nabla }{\overline{P}}+{\rho }_{s}\sigma {\nabla }\overline{T}-{{\boldsymbol{\overline{F}}}}_{{sn}}, \label{eq:TwoFluidLESInviscid} \\ 
{\rho }_{n}\displaystyle \frac{{D}{{\boldsymbol{\overline{v}}}}_{n}}{{D}t}&=& -\displaystyle \frac{{\rho }_{n}}{\rho }{\nabla }{\overline{P}}-{\rho }_{s}\sigma {\nabla }\overline{T}+{\tilde{\eta}}{{\nabla }}^{2}{{\boldsymbol{\overline{v}}}}_{n} - \nabla \cdot \vec{\tau}_{SGS} +{{\boldsymbol{\overline{F}}}}_{{sn}}. \label{eq:TwoFluidLESViscid} 
\end{eqnarray}
Here, $\vec{\tau}_{SGS}$ is the SGS stress, whose components $\tau_{SGS, ij}$ are given as
\begin{eqnarray}
\tau_{SGS, ij} &\equiv& R_{ij} + L_{ij} + C_{ij}, \\
R_{ij} &=& \overline{u_{i}^{\prime} u_{j}^{\prime}}, \\
L_{ij} &=& \overline{\overline{u}_{i} \overline{u}_{j}} - \overline{u_{i}^{\prime} u_{j}^{\prime}}, \\
C_{ij} &=& \overline{u_{i}^{\prime}\overline{u}_{j} + u_{j}^{\prime}\overline{u}_{i}},
\end{eqnarray}
where $\boldsymbol{v}_{n} = (u_{1}, u_{2}, u_{3})^{\rm T}$ and $i,j = 1,2,3$.
As for temperature $T$, we did not solve the heat transport equation by defining the heat flux. Although it may be necessary to do so in future work, we did not consider solving the heat-transport equation at this stage. Therefore, $T$ should be treated as a constant. However, in the SPH calculations, the temperature is obtained by a weighted calculation using the smoothed kernel function with respect to neighboring particles; thus, it varies around the average value depending on the density distribution. 
Therefore, $T = \overline{T} + T_{\epsilon}$ is always true and $T$ can be treated as a single-value function of $\rho$. Briefly, the treatment in the equation is the same as that for $P$. Accordingly, we allow a formal scale transformation of the temperature. In addition, we did not delve into the detailed expression of $\overline{\boldsymbol{F}}_{sn}$.
Our current understanding of the mutual friction force ${\boldsymbol{F}}_{sn}$ is discussed later. 

Equation (\ref{eq:TwoFluidLESViscid}) expresses the Navier--Stokes equations for large-scale viscous fluids, with the subgrid-scale (SGS) stress tensor $(-\nabla \cdot \vec{\tau}_{SGS}$) explicitly representing small-scale effects, formally corresponding to Eq. (\ref{eq:tfmmacrocondiff}). Specifically, the filtered variables $\overline{\boldsymbol{v}}_{n}$, $\overline{\boldsymbol{v}}_{s}$, $\overline{P}$, $\overline{T}$, and $\overline{\boldsymbol{F}}_{sn}$ in Eq. (\ref{eq:TwoFluidLESViscid}) align with their counterparts in Eq. (\ref{eq:tfmmacrocondiff}), and the effective viscosity coefficient $\tilde{\eta}$ satisfies $\tilde{\eta} = \overline{\eta} + \overline{\eta}_r$.
In fact, the rotational viscosity ($\nabla \times \vec{\omega}_{0}$) in Eq. (\ref{eq:tfmmacrocondiff}) represents the parameterization of the internal degrees of freedom of the microscopic molecules to reproduce the spinning motion of the molecules, i.e., the nondissipative vortices. We will discuss later that this term allows the transfer of the motions of the small vortices below the SGS (quantum vortices) to the classical hydrodynamic equation systems at the macroscopic scale. 
Although the divergence of the SGS stress tensor ($-\nabla \cdot \vec{\tau}_{SGS}$) is known to transfer the microscale vortex contributions to the larger scale as mentioned above, the SGS model assumes that the vortices dissipate below the Kolmogorov microscale. Thus, it would be natural to consider the cryogenic liquid helium-4 as an exceptional case where the vortices do not dissipate even below the Kolmogorov microscale due to the loss of viscosity. Before describing this relation, let us consider the Kolmogorov microscale at a high Reynolds number. Several experiments have shown that the viscosity decreases by about 1/8 before and after the critical temperature~\cite{Kapitza1938, BURTON1935}. The Kolmogorov microscale $l$ is given by $l = (\nu^{3}/\epsilon)^{\frac{1}{4}}$, where $\nu$ is the kinematic viscosity coefficient and $\epsilon$ is the mean kinetic energy dissipation rate~\cite{GREGORY200625, STEINBERG2021100900, OKAWA202177}. Thus, when the viscosity is reduced by a factor of 8, $l$ becomes approximately $4.7568\cdots \approx 4.8$ times smaller. The Kolmogorov scale of liquid helium-4 is usually recognized on the scale of from nanometers~\cite{doi:10.1073/pnas.1312546111, Barenghi2016, doi:10.1073/pnas.2018406118} through the intervortex spacing~\cite{babuin:hal-00984313} depending on the conditions. 

\HL{In the following, we show that the Condiff viscosity model can serve as an SGS model and incorporate the quantum vortex interactions into the two-fluid model based on classical hydrodynamics. Recall that the microscale velocity field generated by a quantum vortex filament can be described by the Biot--Savart law. The third and fourth terms on the right-hand side of Eq. (\ref{eq:tfmmacrocondiff}) represent the shear and bulk viscosities, respectively, and the fifth term on the right-hand side of Eq. (\ref{eq:tfmmacrocondiff}) represents the contributions from the rotations of the molecules (or constituent particles) around their axes. We call the fifth term the ``rotational viscosity term.'' Let us decompose this term into the contributions from the respective quantum vortex filaments. To distinguish the difference in spatial scales, the macroscopic version of $\vec{\omega}_{0}(\vec{r}, t)$ in the two-fluid model at the large scale is marked with a superscript line as $\overline{\vec{\omega}_{0}(\vec{r}, t)}$, similar to the cases of Eqs.~(\ref{eq:TwoFluidLESInviscid}) and (\ref{eq:TwoFluidLESViscid}).
In the following, we specify the position $\vec{r}$ and the time $t$ as necessary for easier explanation. The macroscopic variables in Eqs.~(\ref{eq:TwoFluidLESInviscid}) and (\ref{eq:TwoFluidLESViscid}) are obtained by convolving the corresponding microscopic variables with a filter function based on the LES concept. In contrast, for $\vec{\omega}_{0}(\vec{r}, t)$, we admit that $\overline{\vec{\omega}_{0}(\vec{r}, t)}$ can be obtained by the convolution integration of $\vec{\omega}_{0}$ with a distance function $1/|\vec{r}|$ as follows:}
\begin{eqnarray}
\HL{\overline{\vec{\omega}_{0}(\vec{r}, t)} \coloneqq \frac{1}{C_{q}} \int \frac{\vec{\omega}_{0}(\vec{r}', t)}{|\vec{r} - \vec{r}'|} d \vec{r}',} \label{eq:transomega} 
\end{eqnarray}
\HL{where we have introduced a constant coefficient $C_{q}$. Equation~(\ref{eq:transomega}) shows that the microscopic variable $\vec{\omega}_{0}(\vec{r}, t)$ defined on a molecule is transformed into the macroscopic field variable $\overline{\vec{\omega}_{0}(\vec{r}, t)}$ defined around the molecule. The fact that $\overline{\vec{\omega}_{0}(\vec{r}, t)}$ is not defined on a molecular particle at position $\vec{r}$ is consistent with the fluid mechanical picture in that the center of a vortex is a singularity. Although the Gaussian filter is appropriate for the other variables, Eq.~(\ref{eq:transomega}) is more appropriate for $\overline{\vec{\omega}_{0}(\vec{r}, t)}$ because it reproduces the singularity. Let us denote the scale transformation used in Eq.~(\ref{eq:transomega}) as ${\mathcal F[\vec{X}]}$, where $\vec{X}$ is $\vec{\omega}_{0}$ in this case. We also postulate a rigid-body rotational flow; the vorticity $\vec{\omega}$ at any point on a vortex filament can be expressed as $2\vec{\omega}_{0}$. Note that the vorticity of a molecule (or constituent particle) is a different concept from that of a bulk fluid. As explained in Ref.~\cite{10.1063/5.0218444}, the vorticity of a bulk fluid is expressed as ($\nabla \times \vec{v}$), the vorticity due to the internal freedom of the molecule is given by $2\vec{\omega}_{0}$ under the rigid-body rotation assumption, and the difference ($\nabla \times \vec{v} - 2\vec{\omega}_{0}$) is assumed to contribute to the asymmetric part of the stress tensor in the Condiff's model. See Section II.A and Fig. 9 in Ref.~\cite{10.1063/5.0218444} for the details of these relations.} 

\HL{The rotational viscosity term on the right-hand side of Eq. (\ref{eq:tfmmacrocondiff}) can be written using Eq.~(\ref{eq:transomega}) as
\begin{eqnarray}
2 \overline{\eta_{r}} \nabla \times \overline{\vec{\omega}_{0}(\vec{r}, t)} &=& \biggr ( \frac{2 \overline{\eta_{r}}}{C_{q}} \biggl )  \nabla \times \int \frac{\vec{\omega}_{0}(\vec{r}', t)}{|\vec{r} - \vec{r}'|} d\vec{r}' \nonumber \\ 
&=&\biggr ( \frac{2 \overline{\eta_{r}}}{C_{q}} \biggl ) \int \frac{\vec{\omega}_{0}(\vec{r}', t) \times (\vec{r} - \vec{r}')}{|\vec{r} - \vec{r}'|^{3}} d\vec{r}' 
~~+~~\biggr ( \frac{2 \overline{\eta_{r}}}{C_{q}} \biggl ) \int \frac{\nabla \times \vec{\omega}_{0}(\vec{r}', t)}{|\vec{r} - \vec{r}'|} d\vec{r}'. \label{eq:biotsarvvoltwopart}
\end{eqnarray}
\HL{Using ${\mathcal F}[\vec{X}]$, Eq.~(\ref{eq:biotsarvvoltwopart}) can be expressed as follows:}
\begin{eqnarray}
\HL{\therefore~~2 \overline{\eta_{r}} \nabla \times \overline{\vec{\omega}_{0}(\vec{r}, t)}~~=~~\biggr ( \frac{2 \overline{\eta_{r}}}{C_{q}} \biggl ) \int \frac{\vec{\omega}_{0}(\vec{r}', t) \times (\vec{r} - \vec{r}')}{|\vec{r} - \vec{r}'|^{3}} d\vec{r}' 
~~+~~{\mathcal F}[(2 \overline{\eta_{r}} \nabla \times \vec{\omega}_{0}(\vec{r}, t))].} 
\label{eq:biotsarvvolume}
\end{eqnarray}
Here, we have used the relations $\nabla(1/|\vec{r} - \vec{r}'|) = -(\vec{r}-\vec{r}')/|\vec{r} - \vec{r}'|^{3}$ and $\nabla (f \vec{A}) = $ $\nabla f \times \vec{A} + $ $f(\nabla \times \vec{A})$ to obtain Eq.~(\ref{eq:biotsarvvoltwopart}) and Eq.~(\ref{eq:biotsarvvolume}), where $f$ and $\vec{A}$ are the scalar and vector quantities, respectively.}
\HL{The first and second terms on the right-hand side of Eq.~(\ref{eq:biotsarvvolume}) represent the interactions among the vortices and the scale transformation of the rotational viscosity term, respectively. Notably, Eq.~(\ref{eq:biotsarvvolume}) has the recurrence form of $\vec{X}^{(n)} = \vec{B}^{(n+1)} + {\mathcal F}[\vec{X}^{(n+1)}]$, where $\vec{X} = \nabla \times \vec{\omega}_{0}$ and $\vec{B}$ represents the vortex-vortex interactions, with the upper subscript $n$ signifying the depth level of the subgrid scale; thus, the effect of the rotational viscosity terms at the lower hierarchical scales is transferred to the rotational viscosity terms at the upper hierarchical scales. Therefore, the Condiff viscosity model functions as an SGS model.}

\HL{Let us consider the case where the fluid at the subgrid scale follows quantum hydrodynamics, with each quantum vortex having a vorticity $\vec{\omega}$ (= $2\vec{\omega}_{0}$). In this case, the second term on the right-hand side of Eq.~(\ref{eq:biotsarvvolume}) vanishes because the quantum vortices are minimal and there are no smaller-scale vortices. In quantum hydrodynamics, the vorticity $\vec{\omega}$ is localized at the vortex filaments and is thus expressed as}
\begin{eqnarray}
\HL{\vec{\omega}(\vec{r}, t)~=~\kappa \int_{\varGamma} \vec{s}'(\xi, t) \delta (\vec{r} - \vec{s}(\xi, t)) d\xi}, \label{eq:vorticityfilam}
\end{eqnarray}
where \HL{$\varGamma$ represents the line integral along the vortex filaments. $\xi$ is an arc length and $\vec{s}$ is a position vector from the origin to position $\xi$ on the vortex filaments. $\kappa$ represents the quantum of circulation. $\vec{s}'$ is the tangential vector at position $\xi$, and $\vec{s}''$ is perpendicular to $\vec{s}'$; the vectors $\vec{s}'$, $\vec{s}''$, and ($\vec{s}' \times \vec{s}''$) are perpendicular to each other in space and satisfy $\frac{\partial \vec{s}}{\partial \xi} = \vec{s}'$ and $\frac{\partial^{2} \vec{s}}{\partial \xi^{2}} = \vec{s}''$. Using these relations and Eq.~(\ref{eq:vorticityfilam}), we can obtain the following from Eq.~(\ref{eq:biotsarvvolume}):}
\begin{eqnarray}
\HL{\therefore~~2 \overline{\eta_{r}} \nabla \times \overline{\vec{\omega}_0}
~=~\biggr ( \frac{\overline{\eta_{r}}\kappa}{C_{q}} \biggl ) \int_{\varGamma} \frac{\vec{s}'(\xi, t) \times (\vec{r} - \vec{s}(\xi, t))}{|\vec{r} - \vec{s}(\xi, t)|^{3}} d \xi.} \label{eq:biotsavfin}
\end{eqnarray}
\HL{Notably, the right-hand side of Eq.~(\ref{eq:biotsavfin}) corresponds to the Biot--Savart law. In summary, the system of equations for normal fluid components in the two-fluid model based on classical hydrodynamics with the SGS model in Condiff's form, which includes vortex interactions according to the Biot--Savart law, can be expressed as follows:}
\begin{eqnarray}
\HL{\rho_{n} \frac{{D} \overline{\boldsymbol{v}}_{n}}{{D} t} = -\frac{\rho_{n}}{\rho}\nabla \overline{P} - \rho_{s}\sigma\nabla \overline{T} + (\bar{\eta} + \bar{\eta_{r}})\nabla^2 \overline{\boldsymbol{v}}_{n} 
	+ \biggl( \frac{\bar{\eta}}{3} + \bar{\xi} -\bar{\eta_{r}} \biggr) \nabla\nabla\cdot\overline{\boldsymbol{v}}_{n} 
	+ 2\bar{\eta_{r}}\nabla\times \overline{\vec{\omega}_{0}}
	+ \overline{\boldsymbol{F}}_{sn}, } \label{eq:goveqnormal:mutsgs}
\end{eqnarray}
\HL{where the fifth term on the right-hand side of Eq.~(\ref{eq:goveqnormal:mutsgs}) is described as Eq.~(\ref{eq:biotsavfin}).}

Furthermore, we have the following perspective on $\boldsymbol{F}_{sn}$ at the microscopic scale. According to quantum hydrodynamics, the mutual friction force $\boldsymbol{F}_{sn}$ is proportional to the product of the instantaneous local relative velocity $\boldsymbol{v}_{sn}$ and the square of the ensemble average of the relative velocity $\boldsymbol{U}_{sn}$, which is correlated with the statistical average of the vortex line density~\cite{Cooper2019, 10.1063/1.4828892}. This relationship is expressed as follows~\cite{doi:10.1098/rspa.1957.0071, doi:10.1098/rspa.1957.0072, doi:10.1098/rspa.1957.0191, doi:10.1098/rspa.1958.0007}:
\begin{eqnarray}
\boldsymbol{F}_{sn}~~=~~A \rho_{s} \rho_{n} {U}_{sn}^{2} \boldsymbol{v}_{sn}~~=~~\frac{2}{3}\rho_{s}\alpha\kappa L \boldsymbol{v}_{sn}, \label{eq:mutualfric}	
\end{eqnarray}
where ${\rm U}_{sn} = \sqrt{|\boldsymbol{U}_{sn}|^{2}}$ and $A$ is a constant parameter. $\rho_{s}$ and $\rho_{n}$ are the mass densities of the superfluid and normal fluid, respectively. $\alpha$ denotes the mutual friction coefficient. $\kappa$ is the quantum of the circulation. $L$ is the vortex line density, which is obtained as the average length of the vortex configuration over the reduced volume $\Omega$ as follows:
\begin{eqnarray}
L = \frac{1}{\Omega}\int_{\Omega} d\xi, 
\end{eqnarray}
where $\xi$ represents the positions of the vortex lines in $\Omega$. Equation~(\ref{eq:mutualfric}) shows that ${\rm U}_{sn}^2$ is represented as ${\rm U}_{sn}^2 = (\frac{2 \alpha \kappa}{3 A \rho_{n}}) L$, indicating that the substance of ${\rm U}_{sn}^2$ is the spatial average of the vortex line density $L$. 

Accordingly, we can determine that ${U}_{sn}^{2}$ is a macroscopic quantity in quantum hydrodynamics.
If the average to obtain ${U}_{sn}^{2}$ is sufficiently macroscopic to reach the hydrodynamic scale, ${U}_{sn}^{2}$ can be reused in filtering, and only the fluctuation of the relative velocity $\boldsymbol{v}_{sn}$ needs to be considered as $\boldsymbol{F}_{sn} = A \rho_{s} \rho_{n} \overline{U}_{sn}^{2} (\boldsymbol{v}_{sn} + \boldsymbol{v}_{\epsilon})$. However, if the range of velocities for the average is still small compared to the hydrodynamic scale, ${U}_{sn}^{2}$ must also account for the velocity fluctuations as $\boldsymbol{F}_{sn} = A \rho_{s} \rho_{n} (|{\boldsymbol{U}}_{sn} + {\boldsymbol{U}}_{\epsilon}|)^{2} (\boldsymbol{v}_{sn} + \boldsymbol{v}_{\epsilon})$. In the latter case, the nonlinear effects manifest more readily. Future studies should examine the nonlinear effects of $\overline{\boldsymbol{F}}_{sn}$.

\subsection{A proposed view on the composition of liquid helium at microscopic scales}
We explored the possibility that two-fluid models based on classical and quantum hydrodynamics are related via a scale transformation by LES and an inverse scale transformation by SPH in liquid helium-4. We also showed that solving the classical two-fluid model using SPH allows the smoothing kernel approximation errors to mimic microscopic variations, effectively reproducing them at macroscopic scales depending on the kernel radius $h$. We now focus on the spatial distribution of inviscid versus viscous components, and superfluid versus normal fluid components, and examine their interrelation.
Landau's original two-fluid model assumed that superfluid and normal fluid components coexist independently within the same volume. This view has since evolved. In recent quantum hydrodynamic models, only the normal fluid---representing the averaged flow of quantum excitations---occupies the entire space. Although normal fluids are influenced by quantum vortex tangles, these vortices appear as singularities on the macroscopic scale and thus contribute no volume. While normal fluid is typically regarded as composed solely of viscous fluid particles, this may be an oversimplification. In macroscopic quantum phenomena, such as in liquid helium-4, the overall fluid viscosity decreases significantly, suggesting that the bulk behaves as an inviscid fluid. This contradicts the current quantum two-fluid model, where the entire domain is considered filled with normal fluid. Accordingly, a revised correspondence among the concepts of inviscid fluid, viscous fluid, superfluid, and normal fluid is needed.

\begin{figure}[t]
\vspace{-30.0cm}
\hspace{26.5cm}
\centerline{\includegraphics[width=4.2\textwidth, clip, bb= 0 0 4223 2265 ]{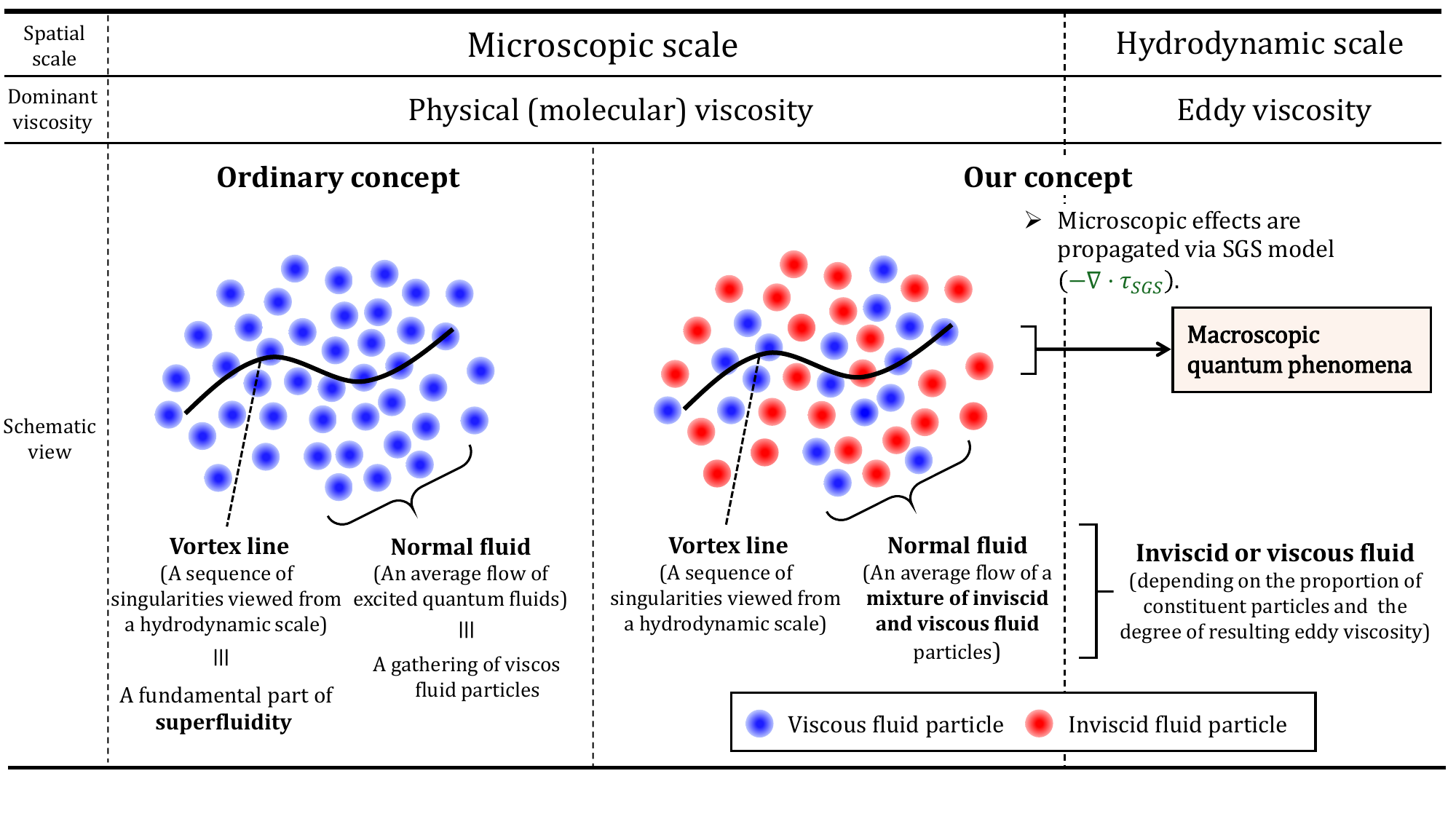}}
\caption{A proposed view of the microscopic composition of cryogenic liquid helium-4 in a multiscale framework.}
\label{fig:Figure-ClassificationConcept}
\end{figure}

Figure~\ref{fig:Figure-ClassificationConcept} illustrates the proposed microscopic composition of cryogenic liquid helium-4 within a multiscale framework. At finite cryogenic temperatures, the fluid can be modeled as an average flow of inviscid and viscous fluid particles. The presence of viscous particles causes the system to behave as a normal fluid at microscopic scales. However, while molecular viscosity governs this local behavior, its small magnitude contributes minimally to the large-scale effective viscosity, which also includes eddy viscosity. Thus, in laminar regimes where eddy viscosity is negligible, the normal fluid may be treated as effectively inviscid at macroscopic scales if molecular viscosity is sufficiently low.This view contrasts with recent models that assume the entire system is filled with a normal (viscous) fluid at macroscopic scales. It is instead more consistent with Landau's original two-fluid model, which allows for the coexistence of inviscid and viscous components across spatial regions. Notably, the counterflow simulations in Fig.~\ref{fig:Figure-ResultsCounterFlow}, which matched the spatial distribution of the normal fluid with the vortex line density, successfully reproduced the tail-flattened velocity profile. This suggests that residual viscous fluid particles at finite temperatures may act as seeds for quantum vortex lines through local field ionization. Supporting this, helium-3 impurities and ions~\cite{PhysRevB.16.244, doi:10.1098/rspa.1985.0116} in liquid helium-4 are known to generate vortex rings~\cite{PhysRev.136.A1194, doi:10.1098/rsta.1985.0003} and loops~\cite{doi:10.1098/rsta.1984.0038, doi:10.1098/rsta.1990.0122}. From a classical standpoint, vortices are low-density regions. Since normal fluid components at cryogenic temperatures are also low in density, as predicted by BEC theory, they are plausible candidates for vortex cores. In summary: (1) the normal fluid may comprise both inviscid and viscous fluid particles; (2) although molecular viscosity governs microscopic behavior, its small magnitude has little effect on large-scale viscosity, allowing the normal fluid to be treated as inviscid in laminar regimes; and (3) at finite temperatures, some viscous fluid particles may serve as seeds for quantum vortices via field ionization.

\section{Conclusion}
Our recent numerical studies on cryogenic liquid helium-4 highlight key features of multiscale physics that can be captured using the two-fluid model. In this paper, we demonstrated that classical and quantum hydrodynamic two-fluid models are connected via scale transformations: large eddy simulation (LES) filtering links micro to macroscales, while inverse scale transformation through SPH connects macro back to microscales. We showed that the spin angular momentum conservation term formally corresponds to a subgrid-scale (SGS) model derived from this transformation. 
\HLLL{Moreover, solving the classical hydrodynamic two-fluid model with SPH appears to reproduce microscopic-scale fluctuations at macroscopic scales.} 
\HLLL{In particular,} the amplitude of these fluctuations depends on the kernel radius. 
\HLLL{This effect may be attributed to truncation errors from kernel smoothing, which can qualitatively resemble such fluctuations.} 
\HLLL{However, this resemblance lacks first-principle justification and should be viewed as a speculative analogy rather than a physically grounded effect.}
Our theoretical analysis further suggests that the Condiff viscosity model can act as an SGS model, incorporating quantum vortex interactions under point-vortex approximation into the two-fluid framework. These findings provide new insight into the microscopic structure of cryogenic helium-4 within a multiscale context. Specifically: (1) the normal fluid can be understood as a mixture of inviscid and viscous particles; and (2) \HLL{while molecular viscosity renders the normal fluid at microscopic scales, its small magnitude contributes little to the large-scale effective viscosity, which includes both molecular and eddy viscosities; therefore, in laminar regimes where eddy viscosity is negligible, the normal fluid may be effectively treated as inviscid at large scales if molecular viscosity is sufficiently small.} In conclusion, we have established a multiscale physical framework for modeling cryogenic liquid helium-4 that bridges classical and quantum hydrodynamic regimes.

\section*{Acknowledgment}
This study was supported by JSPS KAKENHI Grant Number 22K14177 and JST PRESTO, Grant Number JPMJPR23O7.
The authors thank Editage (www.editage.jp) for the English language editing.
The author would also like to express gratitude to his family for their moral support and encouragement.

\bibliographystyle{elsarticle-num}
\bibliography{reference}

\end{document}